\documentclass[12pt]{article}
\usepackage{latexsym,amsmath,amssymb,epsfig,graphicx}

\newcommand{\pkt}{\; .}
\newcommand{\kma}{\; ,}
\newcommand{\nn}{\nonumber}
\newcommand{\bea}{\begin{eqnarray}}
\newcommand{\eea}{\end{eqnarray}}
\newcommand{\be}{\begin{equation}}
\newcommand{\ee}{\end{equation}}
\newcommand{\beast}{\begin{eqnarray*}}
\newcommand{\eeast}{\end{eqnarray*}}
\newcommand{\eqn}[1]{(\ref{#1})}
\newcommand{\cald}{{\cal D}}
\newcommand{\cali}{{\cal I}}
\newcommand{\calk}{{\cal K}}
\newcommand{\call}{{\cal L}}
\newcommand{\calg}{{\cal G}}
\newcommand{\calm}{{\cal M}}
\newcommand{\caln}{{\cal N}}
\newcommand{\calf}{{\cal F}}
\newcommand{\calu}{{\cal U}}
\newcommand{\calb}{{\cal B}}
\newcommand{\calo}{{\cal O}}
\newcommand{\bfm}{{\bf M}}

\begin{document}
%
%%%%%%%%%%%%%%%%%%%%%%%%%%%%%%%%%%%%%%%%%%%%%%%%%%%%%%%%%%%%%%%%%%%%%%%%%
\begin{titlepage}
\begin{flushright}
DO-TH-06/11 \\
hep-th/0611004 \\
October 2006
\end{flushright}

\vspace{20mm}
\begin{center}
{\Large \bf
False vacuum decay by self-consistent bounces in four dimensions}
\vspace{10mm}

{\large
J\"urgen Baacke $^a$ \footnote{e-mail: baacke@physik.uni-dortmund.de}
and
Nina Kevlishvili $^{a,b}$  
\footnote{e-mail: nina.kevlishvili@het.physik.uni-dortmund.de}}

\vspace{15mm}

{\large  $^a$  Institut f\"ur Physik, Universit\"at Dortmund \\
D - 44221 Dortmund, Germany\\
and\\
$^b$ Andronikashvili Institute of Physics,\\ GAS, 0177 Tbilisi, Georgia
}

\vspace{15mm}

\bf{Abstract}
\end{center}
We compute bounce solutions describing false vacuum decay
in a $\Phi^4$ model in four dimensions with quantum back-reaction.
The back-reaction of the quantum fluctuations on the 
bounce profiles is computed in the one-loop and  Hartree approximations.
This is to be compared with the usual semiclassical approach where
one computes the profile from the classical action and determines
the one-loop correction from this profile. The computation of the 
fluctuation determinant is performed using a theorem on functional 
determinants,
in addition we here need the Green' s function of the fluctuation operator
in oder to compute the quantum back-reaction.
As we are able to separate from the determinant and from the Green' s function
the leading perturbative orders, we can regularize and renormalize
analytically, in analogy of standard perturbation theory.
The iteration towards self-consistent solutions is found to converge 
for some range of the parameters. Within this range the corrections to
the semiclassical action are at most a few percent, the corrections
to the transition rate can amount to several orders of magnitude. 
The strongest 
deviations happen for large couplings, as to be expected.
Beyond some limit, there are no self-consistent bounce solutions.

\end{titlepage}

\setcounter{page}{2}

\section{Introduction}\label{introduction}
\setcounter{equation}{0}

False vacuum decay \cite{Coleman:1977py,Callan:1977pt,Linde:1981zj}
is one of the basic mechanisms which are often
invoked in the construction of cosmological models or scenarios.
False vacuum decay may initiate inflation, or it may happen
after inflation, if the universe gets trapped in one of the
local minima of the Higgs potential in Grand Unified Theories.
The false vacuum may decay by spinodal decomposition
\cite{Cormier:1999ia,Cormier:1998nt,Boyanovsky:1999wd} if the
minimum becomes a maximum by a change of the effective potential,
e.g., with decreasing temperature or cosmological expansion,
or triggered by another field (inflaton)\cite{Garcia-Bellido:2002aj}. 
If this does not happen
it may decay by an over-the-barrier transition at finite temperature,
the process of bubble nucleation, or it may tunnel to the
true vacuum, a process that is described semiclassically by
bounce solutions of the associated Euclidean theory.
If gravity is included, such a solution is the
Coleman-de Luccia bounce \cite{Coleman:1980aw,Hackworth:2004xb}. 
We will concentrate here on
bounce solutions in a theory with one single scalar field
and an asymmetric double-well potential. The methods described here
may be transferred to more realistic models. In particular, they
may also be used  for the multifield
case with coupled equations for the classical and fluctuations fields.

The most common approach to computing the tunneling rate
is the semiclassical one. The bounce profile
is determined by minimizing the classical Euclidean action while 
in the calculation of the transition rate one includes the one-loop 
prefactor, i.e., the fluctuation determinant computed
around the classical profile. For false vacuum decay
in $3+1$ dimensions this computation to one-loop accuracy
has been presented in Ref. \cite{Baacke:2003uw}, based on methods
developed in \cite{Baacke:1993ne}. Recently
similar computations have been
presented in Ref. \cite{Dunne:2005rt} using $\zeta$ function
regularization. A different technique for computing the
fluctuation determinant has been used in 
\cite{Baacke:1994bk,Baacke:1994ix,Baacke:1993aj}.

Once the quantum corrections become important one may ask whether
these quantum fluctuations react back upon the bounce profile.
Two ways of including the quantum backreaction for such semiclassical
systems have been discussed in the literature:  the one-loop 
backreaction and the Hartree backreaction. 
The backreaction to  one-loop order  consists in determining 
the classical field
in such a way that it is an extremum of the one-loop effective action,
the sum of classical and one-loop action. 
This has been discussed by Surig \cite{Surig:1997ne}
 in the context of bubble nucleation in a $SU(2)$ gauge theory
(a multifield case). We will call this the one-loop backreaction.
 In the Hartree approximation one takes into 
account the backreaction not only upon the classical field, but also 
upon the quantum fluctuations. We will call this the Hartree
backreaction. This approach has been applied
recently, using quite different mathematical and numerical approaches,
by Bergner and Bettencourt \cite{Bergner:2003id} and by the present authors
\cite{Baacke:2004xk} to the case of false vacuum decay in two space-time
dimensions. Here we extend the latter scheme to $3+1$ space-time dimensions,
i.e. four Euclidean dimensions for the bounce. We also present results 
for the one-loop backreaction. 

The methods used for the one-loop and the Hartree approaches are 
closely related.
In both cases one has to compute not only the fluctuation determinant
but also  the Green function $\calg(x,x')$ of the fluctuation
operator. This is done in analogy to the two-dimensional case
\cite{Baacke:2004xk}.
For the case of a bounce in four Euclidean dimensions 
renormalization becomes
more involved than the one in two dimensions, by the occurrence
of logarithmic, quadratic and quartic divergences, both in the Green's
function at $x=x'$ and in the fluctuation determinant.
 So regularization and renormalization
have to be discussed more extensively. Besides the
regularization technique we will have to discuss the 
renormalization conditions. One way of renormalizing is to
use the $\overline{MS}$ prescription on the basis
of dimensional regularization. This implies a rather transparent
choice of counterterms and it would
be appropriate if the bounce were computed for a theory, in which
the parameters are determined from experiment, using the same prescription.
This prescription has been  used in  Refs.
\cite{Baacke:2003uw,Baacke:2004xk,Dunne:2005rt}.
On the other hand, in our toy world, one may wish to compare the results
of the various approaches for situations were
the effective potentials have similar "physical" properties, e.g. 
 the same position of the local minima and the
same energy differences between the two vacua. This approach is intuitively
more appealing, while the determination of the counterterms 
gets somewhat less transparent. These renormalization conditions were used
in Ref.\cite{Baacke:1993ne}. In the latter publication the authors
have used Pauli-Villars regularization. This could have been done here as
well; as the divergent parts are separated in a Lorentz-invariant
way from the finite parts which are computed numerically, 
the method of regularization can be chosen freely, as in usual perturbative
calculations.

The plan of the paper is as follows:
In section \ref{basicformulae} we present the basic equations of the 
theoretical model and for the tunneling rate. In section \ref{Hartreebounce}
we introduce the formalism used to compute the quantum backreaction
in the Hartree approximation, using the $2PPI$ expansion for the
effective action. In section
\ref{oneloopbounce} we present the formalism for computing the 
quantum backreaction to one-loop accuracy, as the lowest order of the
$1PI$ action. 
  We then turn to the analytical and numerical methods.
 In sections \ref{greensfunction} and \ref{flucdet} we introduce 
the mathematical formalism
used for the numerical computation of the Green's function
and of the fluctuation determinant, respectively. 
In section \ref{unstableandtranslation} we discuss the 
particularities associated with the
unstable mode in the $n=0$ and with the translation mode in the
$n=1$ partial waves, and the  numerical techniques used to overcome
these problems. The renormalization of the dynamics is discussed
in section \ref{renormalization}, the technical
details are deferred to Appendices \ref{appendix:effpot} and 
\ref{appendix:renormaction}.
The numerical results are presented and discussed in section
\ref{numericalresults}. We summarize our approach and the results
in section \ref{summary}.

%%%%%%%%%%%%%%%%%%%%%%%%%%%%%%%%%%%%%%%%%%%%%%%%%%%%%%%%%%%%%%%%%%%%%%%%%

\section{Basic relations} \label{basicformulae}
\setcounter{equation}{0}
\label{thebounce}
We consider a scalar field theory in $3+1$ dimensions, with the Lagrange
density
\be
\call=\frac{1}{2} \partial_\mu \Phi \partial^\mu \Phi -U(\Phi)
\pkt\ee
The  potential $U(\Phi)$ is given by
\begin{equation}
U(\Phi)=\frac{1}{2} m^2\Phi^2-\eta\Phi^3
+\frac{1}{8}\lambda\Phi^4 \kma
\end{equation}
and is represented in Fig. \ref{figure:potential} (solid line).
It displays two minima, one at $\Phi=0$, representing the false vacuum,
and the other one at $\Phi=\phi_{\rm tv} > 0$ corresponding to the true vacuum.

The classical Euclidean action of this theory is given by
\be
S_{\rm cl}[\phi]=\int d^4 x \left[\frac{1}{2}\left(\nabla \phi\right)^2
+U(\phi)\right]\kma
\ee
and the bounce which minimizes this action satisfies
\be
\left[-\left(\frac{\partial}{\partial t}\right)^2-\Delta\right]\phi 
+U'(\phi)=0\kma
\ee
or, using its spherical symmetry,
\be
-\frac{d^2\phi}{dr^2}-\frac{3}{r}\frac{d\phi}{dr} + U'(\phi)=0\pkt
\ee
The boundary conditions are
\begin{equation}\label{bounceboundaryconditions}
\frac{d\phi}{dr}|_{r=0}=0,\qquad \phi_{r\to\infty}=\phi_{\rm tv}\pkt
\end{equation}

The one-loop correction to the classical action is given by
\be
S_{\rm 1-l}=\frac{1}{2}\ln {\det}' \frac{-\partial^2+U''(\phi( x))}
{-\partial^2+m^2}=\frac{1}{2}\ln \cald[\phi]\kma
\ee
where $\partial^2$ is the Laplace operator in four dimensions,
$m$ is the mass in the false vacuum,
\be
m^2=U''(0)
\pkt
\ee
and where the prime indicates that the translation zero mode is removed
and that one replaces the imaginary frequency of the unstable mode
by its absolute value.

The transition rate from the false to the true vacuum  is given,
in the semiclassical approximation without backreaction, by
\be \label{transitionrate}
\Gamma^{1-\rm{loop}}=\left(\frac{S_{\rm cl}}{2\pi}\right)^2\cald^{-1/2}
\exp(-S_{\rm cl})
=\left(\frac{S_{\rm cl}}{2\pi}\right)^2\exp(-S_{\rm cl}-S_{\rm 1-l})
\kma\ee
where
\be
\cald[\phi]={\det}'\frac{-\partial^2+\calm^2}{-\partial^2+m^2}
\pkt\ee

We recall (see e.g. \cite{Coleman85}) that the 
prefactor arises from the quantization of the collective 
coordinate associated with the translation zero mode.
The presence of a zero mode $\eta_0$ in this approximation 
is demonstrated by taking the gradient of
the classical equation of motion: 
\be
\nabla_i \left[-\partial^2\phi +U'(\phi)\right]=
\left[-\partial^2 +U''(\phi)\right]\nabla_i \phi=0\pkt
\ee
Its normalization, defined by $\eta_{0i}=\caln_0\nabla_i\phi$,
and the condition that $\eta_{0i}$ is normalized to unity,
is given by
\be
\caln_0^{-2}=\int d^4 x \left(\nabla_i\phi\right)^2=T/2
\kma\ee 
where there is no summation over $i$ and where $T$ is the kinetic
part of the action. Quantization of the collective coordinates
yields the prefactor $(\caln_0^{-2}/2\pi)^2=(T/4\pi)^2$. 
Furthermore, one finds  in 
four dimensions, using a scaling argument, that 
$V=\int d^4x U(\phi)=-T/2$. We then obtain $T=2S_{\rm cl}$ and 
the prefactor becomes
$(T/4\pi)^2=(S_{\rm cl}/2\pi)^2$, as written 
down in Eq. \eqn{transitionrate}.

%%%%%%%%%%%%%%%%%%%%%%%%%%%%%%%%%%%%%%%%%%%%%%%%%%%%%%%%%%%%%%%%%%%%%%%%%

\section{The bounce in the Hartree approximation}
\setcounter{equation}{0}
\label{Hartreebounce}

The Hartree approximation can be derived in various ways,
in the most intuitive approach from a variational
principle using  a wave function which is a direct
product. Within the so-called 2PPI formalism 
\cite{Coppens:1993zc,Verschelde:1992bs,Verschelde:2000dz}
it is the lowest quantum approximation. In terms of Feynman graphs
it represents a resummation of daisy and super-daisy diagrams, like
in the large-$N$ effective action.
The effective action in this formalism is given by
\be\label{effacH}
S_{\rm eff}[\calm^2,\phi]=
S_{\rm cl}[\phi]+\Gamma^{2PPI}[\calm^2,\phi]
-\frac{3\lambda}{8}\int d^4x \Delta^2( x)
\kma\ee
up to renormalization counterterms discussed in section \ref{renormalization}.
Here $\Delta$ is  a local insertion into the
propagator which has the form
\be
\calg^{-1}( x)=-\partial^2 +\calm^2( x)\kma
\ee
with the definition
\be \label{calmdefH}
\calm^2=m^2-6\eta \phi+\frac{3}{2}\lambda\phi^2
+\frac{3}{2}\lambda \Delta=U''(\phi)+\frac{3}{2}\lambda\Delta
\pkt \ee 
$\Delta$ itself is defined by the  equation
\be
\frac{1}{2}\Delta( x)=\frac{\delta}{\delta \calm^2(x)}
\Gamma^{2PPI}[\phi,\calm^2] 
\pkt \ee
With this definition Eq. \eqn{calmdefH} becomes a
self-consistent equation, the gap equation.
Finally $\Gamma^{2PPI}$ is the sum of all two-particle 
point-irreducible graphs, in which all internal propagators
have the effective masses $\calm^2$. 
A graph is two particle point reducible (2PPR)
if it falls apart when {\em two lines meeting at a point} are cut.
To lowest order in a  loop expansion $\Gamma^{2PPI}$
is given by a simple loop, i.e.
\be \label{defgamma2ppi}
\Gamma^{2PPI}=\frac{1}{2}\ln{\det}'\frac{-\partial^2+\calm^2}
{-\partial^2+m^2}
\kma\ee
and this is equivalwent to the Hartree approximation.
We will discuss the modifications arising from the zero
and unstable modes later.
In this lowest approximation, $\Delta$ is given by
\be\label{calfdef}
\Delta(x)=2\frac{\delta \Gamma^{2PPI}}{\delta \calm^2( x)}=
< x|\frac{1}{-\partial^2+\calm^2}| x>=\calg(x, x)
\equiv \calf( x)\pkt \ee
Here the Green's function $\calg$ is defined by
\be
(-\partial^2+\calm^2)\calg(x,y)=\delta^4(x- y)
\kma\ee
and where we have introduced the fluctuation integral
\be
\calf(x)=\calg( x, x)
\pkt\ee
In taking variational derivatives of the effective action
we have to consider $\Delta$ as a functional of $\calm^2$
and $\phi$, i.e., in the last term
of Eq. \eqn{effacH} we have to replace
\be
\Delta=-\phi^2+\frac{2}{3\lambda}\left(\calm^2-m^2+6\eta\phi\right)
=\frac{2}{3\lambda}(\calm^2-U''(\phi))\kma
\ee
see Eq. \eqn{calmdefH}.

Taking the variational derivative of the effective action
with respect to $\calm^2$ leads back to Eq. \eqn{calmdefH}.
The partial derivative
with respect to the field $\phi$ leads to
the equation for the bounce profile
\be
-\partial^2\phi
+U'(\phi( x))+ \frac{3}{2}
\left[\lambda \phi( x)-2\eta\right]\calf( x)=0
\pkt
\ee
It is not a simple differential or integro-differential equation 
as $\calf(x)$ is a nonlinear {\em functional} of $\phi$. 

Using rotational symmetry we obtain explicitly
\bea\nn
&&-\frac{d^2\phi(r)}{dr^2}-\frac{3}{r}\frac{d\phi(r)}{dr}
+m^2\phi(r)-3 \eta\phi^2(r)+
\frac{\lambda}{2}\phi^3(r) 
\\\nn
&&\hspace{10mm}+ \frac{3}{2}
\left[\lambda \phi(r)-2\eta\right]
\calf( x)\biggl|_{|x|=r}=0
\pkt
\eea
The backreaction of the quantum modes upon themselves is contained
in $\calf( x)$, or, equivalently, 
in $\calm^2( x)$.
For a spherical background field, as we have it here, these functions are
themselves spherically symmetric.

Of course $\calf( x)$ as the limit $ x'\to  x$ 
of the Green's function $\calg( x, x')$ is ill-defined,
as is the fluctuation determinant in Eq. \eqn{defgamma2ppi}.
This problem will be dicussed in section \ref{renormalization}.

%%%%%%%%%%%%%%%%%%%%%%%%%%%%%%%%%%%%%%%%%%%%%%%%%%%%%%%%%%%%%%%%%%%%%%%%%

\section{The bounce with one-loop backreaction}
\setcounter{equation}{0}
\label{oneloopbounce}

The one-loop backreaction can be derived using the $1PI$ formalism
and represents in fact a simplification in comparison with the
Hartree backreaction. The action is given by
\be\label{effac1l}
S_{\rm eff, 1-l}[\calm^2,\phi]=
S_{\rm cl}[\phi]+\Gamma^{1PI}[\phi]
\kma\ee
up to renormalization counterterms. Here $\Gamma^{1PI}[\phi]$
is the sum of all one-particle irreducible graphs with the propagators  
\be
\calg^{-1}( x)=-\partial^2 +\calm^2( x)\kma
\ee
where here $\calm^2$ is given by
\be \label{calmdef1l}
\calm^2=m^2-6\eta \phi+\frac{3}{2}\lambda\phi^2=U''(\phi)
\pkt \ee 
To the lowest order $\Gamma^{1PI}(\phi)$ is given by
\be
\Gamma^{1PI}=\frac{1}{2}\ln{\det}'\frac{-\partial^2+\calm^2}
{-\partial^2+m^2}
\pkt\ee
Taking the  partial derivative
with respect to the field $\phi$ leads to
the equation for the bounce profile
\be
-\partial^2\phi
+U'(\phi( x))+ \frac{3}{2}
\left[\lambda \phi( x)-2\eta\right]\calf( x)=0
\pkt\ee
The last term in this equation arises by
\be
\frac{\delta \Gamma^{1PI}}{\delta\phi( x)}=
\frac{\delta \Gamma^{1PI}}{\delta\calm^2( x)}
\frac{d \calm^2( x)}{d \phi(x)}=
\frac{1}{2}\calf( x)(3\lambda \phi(x)-6\eta)
\kma\ee
where again $\calf( x)=\calg( x, x)$.
The similarities with the Hartree backreaction are obvious,
we will need analogous numerical techniques, the main difference are 
in the effective mass, which in the one-loop formalism is simply
a function of $\phi$, whereas in the Hartree approximation
 it is determined in a self-consistent way.

%%%%%%%%%%%%%%%%%%%%%%%%%%%%%%%%%%%%%%%%%%%%%%%%%%%%%%%%%%%%%%

\section{Computation of the Green's Function}
\label{greensfunction}
\setcounter{equation}{0}

In order to include the backreaction of the quantum fluctuations
on the bounce in the Hartree approximation we need
the Green's function $\calg( x, x')$ of the
(new) fluctuation operator. In fact the Green's function
is usually discussed in a more general form, as a function
of energy. We will need to take this more general
approach in order to be able to discuss the translation mode and in order to 
discuss the determinant theorem in the next section.
Such a concept corresponds to introducing
an additional fifth dimension. We will choose it spacelike,
thus introducing an additional Euclidean time.
As $\phi$ still lives in four Euclidean dimensions, we have translation
invariance in the new time direction.  So we can introduce the Fourier
transform $\calg( x, x',\nu^2)$, where $\nu$ is
the Euclidean frequency. 
The Green's function now satisfies
\begin{equation}
[-\partial^2+m^2+V(r)+\nu^2]\calg(x,x',\nu^2)=\delta^4( x - x')
 \kma
\end{equation}
with
\be
V(r)=\calm^2(\phi)-m^2
\pkt\ee
Here $\calm^2(\phi)$ is given either by Eq. \eqn{calmdefH} or
by Eq.  \eqn{calmdef1l} in the Hartree and in the one-loop formalism,
respectively.
In the one-loop formalism $\calm^2(0)=m^2$, so with the boundary
conditions \eqn{bounceboundaryconditions} $V(r)\to 0$ as
$r\to \infty$. In the Hartree formalism 
$\calm^2$ is, via the Green function, a {\em functional} of $\phi$. 
Here $\calm^2$ should be equal to $m^2$ for a constant field
$\phi(x)\equiv 0$. This has to be imposed as a renormalization
condition. Alternatively, in the definition of $V(r)$ 
one replaces the bare mass $m^2$ with the
renormalized mass in the false vacuum.

The Green's function can be expressed by the eigenfunctions
of the fluctuation operator. We denote them by $\psi_\alpha( x)$,
they satisfy
\be
[-\partial^2+m^2+V(r)]\psi_\alpha( x)=
\omega_\alpha^2 \psi_\alpha( x)
\pkt\ee
In terms of these eigenfunctions the Green's function can be written as
\be\label{greenformal1}
\calg( x,x',\nu^2)=
\sum_\alpha\frac{\psi_\alpha( x)\psi_\alpha(x')}
{\omega_\alpha^2+\nu^2}
\pkt\ee
We may, furthermore, decompose the Hilbert space into angular momentum
subspaces, introducing eigenfunctions $Y_{nlm}(\Omega_3) R^\alpha_k(r)$,
where $Y_{nlm}(\Omega_3)$ are the spherical functions on the
$3$-sphere (see e.g. the Appendix of Ref. \cite{mottola}) and where the 
radial wave functions $R_{n\alpha}(r)$ are eigenfunctions of the partial 
wave fluctuation operator:
\be
\left[-\frac{d^2}{dr^2}-\frac{3}{r}\frac{d}{dr}+\frac{n(n+2)}{r^2}+
m^2+V(r)\right]R_{n\alpha}(r)=\omega_{n\alpha}^2 R_{n\alpha}(r)
\pkt\ee
The index $\alpha$ labels the radial excitations, the spectrum 
is continuous, but may include some discrete states,
e.g., the unstable and translation modes. In terms of these eigenfunctions
the Green's  function takes the form
\be\label{greenformal2}
\calg( x, x',\nu^2)=\sum_{nlm}\sum_\alpha 
Y_{nlm}(\Omega_3)Y_{nlm}(\Omega_3')
\frac{R_{n\alpha}(r)R^*_{n\alpha}(r')}{\omega_{n\alpha}^2+\nu^2}
\pkt \ee
While these expressions are very suitable
for discussions on the formal level, they are not very suitable for numerical
computation. In particular, if one uses these expressions for
the numerical computation, it becomes necessary to discretize
the continuum states by introducing a finite spatial boundary.

There is a well known alternative way of expressing Green's functions.
Consider first the free Green's function obtained for $V(r)=0$.
It can be written as
\be
G_0(x, x',\nu^2)=\int{\frac{d^4k}{(2\pi)^4}\frac{e^{i k\cdot
( x- x')}}{k^2+m^2+\nu^2}}\kma
\ee
and this may be expanded as 
\begin{eqnarray}\nonumber
&&\int\frac{d^4k}{(2\pi)^4}
\frac{e^{ik\cdot (x- x')}}{{ k}^2+m^2+\nu^2}
=\frac{\kappa^2}{4\pi^2}
\frac{K_1(\kappa R)}{\kappa R}\\
&&=\frac{1}{2\pi^2}\sum_{n=0}^\infty(n+1)C_n^1(\cos\chi)
\frac{I_{n+1}(\kappa r_<)}{r_<}\frac{K_{n+1}(\kappa r_>)}{r_>}\nonumber\\
&&=\frac{1}{2\pi^2}\sum_{n=0}^\infty(n+1)
C_n^1(\cos\chi)\cali_{n-}^{(0)}(r_<,\kappa)
\calk_{n+}^{(0)}(r_>,\kappa)\nonumber
\pkt\end{eqnarray}
With $r=| x|$ and $r'=| x'|$ we have
$r_< = \min \{r,r'\}$, $r_> = \max \{r,r'\}$, $\kappa$ is
defined as $\kappa=\sqrt{m^2+\nu^2}$. 
$R$ and $\chi$ are defined by
\be
R^2=|x- x'|^2=r^2+r'^2-2rr'\cos\chi
\kma \ee
i.e., $\chi$ is the angle between the directions  $\Omega_3$ and
$\Omega_3'$ of
$ x$ and $ x'$.
The functions $C_n^1$ are Gegenbauer polynomials, see section
 of Ref. \cite{bateman}. 
The expansion of $K_1(\kappa R)$
in terms of products of $I_l(\kappa r_<)$ and $K_l(\kappa r_>)$
is the Gegenbauer expansion, given in section 7.61, Eq. (3)  
of Ref.\cite{bateman}. 
For the case $x=x'$ one has 
$C_n^1(\cos\chi)=C_n^1(1)=n+1$.
For convenience we have introduced the functions $\cali_n$ and $\calk_n$ as
\bea \nonumber
\cali_n(r,\kappa)&=&I_{n+1}(\kappa r)/r\kma\\
\calk_n(r,\kappa)&=& K_{n+1}(\kappa r)/r
\pkt
\eea
 They satisfy
\be
\left[-\frac{d^2}{dr^2}-\frac{3}{r}\frac{d}{dr}+\frac{n(n+2)}{r^2}+
\kappa^2\right]\calb_n(r,\kappa)=0\kma
\ee
where $\calb_n$ stands for $\cali_n$ or $\calk_n$. Their Wronskian is given by
\be \nonumber
\calk_{n}(r,\kappa)d\cali_{n}(r,\kappa)/dr
-\cali_{n}(\kappa, r)d\calk_{n}(r,\kappa)/dr=1/r^3
\pkt\ee
We expand the {\em exact} Green's function in an analogous way 
with  the ansatz
\be \label{Green}
\calg( x, x',\nu^2)=\frac{1}{2\pi^2}\sum_{n=0}^\infty 
(n+1)C_n^1(\cos\chi)f_n^-(r_<,\kappa)f_n^+(r_>,\kappa)
\pkt\ee
The functions $f_n^{\pm}(r,\kappa)$ 
satisfy the mode equations
\be
\left[-\frac{d^2}{dr^2}-\frac{3}{r}\frac{d}{dr}+\frac{n(n+2)}{r^2}+
\kappa^2+V(r)\right]f_n^{\pm}(r,\kappa)=0\kma
\ee
and the following boundary conditions:
\be
\begin{array}{ll}
f_n^-(r,\kappa)\propto r^n & r\to 0\kma
\\
f_n^+(r,\kappa)\propto \exp(-\kappa r)/\sqrt{\kappa r^3} & r \to \infty
\pkt\end{array}
\ee
So $f_n^-$ is regular at $r=0$ and $f_n^+$ is
regular, i.e., bounded, as $r\to \infty$.
For $V(r)=0$ these boundary conditions are those satisfied by
$\cali_n(r,\kappa)$ and $\calk_n(r,\kappa)$, respectively.  
Furthermore, as the behaviour
at $r=0$ is determined by the centrifugal barrier, and
the behaviour for $r\to \infty$ by the mass term, these boundary conditions
are independent of the potential. If we write
\bea
f_n^-(r,\kappa)&=&\cali_n(r,\kappa)[1+h_n^-(r,\kappa)]\kma
\\
f_n^+(r,\kappa)&=&\calk_n(r,\kappa)[1+h_n^+(r,\kappa)]\kma
\eea
then the functions $h^\pm_n(r,\kappa)$ become constant as $r\to 0$ and 
as $r\to \infty$, and for finite $r$ they interpolate smoothly
between these asymptotic constants.
If we impose, for $r \to \infty$ the boundary 
conditions $h^\pm(r,\kappa)\to 0$ 
the Wronskian of $f_n^+$ and 
$f_n^-$ becomes identical to the one between $K_{n}(\kappa r)$ and
$I_{n}(\kappa r)$, i.e.,  equal to $1/r^3$.
Applying the fluctuation operator to our ansatz, Eq. \eqn{Green},
we then find
\begin{eqnarray}
&&\left[-\partial^2+\kappa^2+V(r)\right]G( x, x',\nu^2)=
\frac{1}{2\pi^2}
\frac{1}{r^3}\delta(r-r')\sum_{n=0}^\infty (n+1)C_n^1(\cos\chi)\nonumber\\
&&=\frac{1}{r^3}\delta(r-r')\frac{1}{\sin^2\chi}\delta(\chi-\chi')
\frac{1}{\sin\theta}\delta(\theta-\theta')\delta(\varphi-\varphi')\kma
\end{eqnarray}
where we have used the addition theorem
\be
\sum_{l=0}^n\sum_{m=-l}^lY_{nlm}(\Omega_3)Y_{nlm}^*(\Omega_3')
=\frac{n+1}{2\pi^2}C_n^1(\cos\chi)
\ee
and the completeness relation for the $O(4)$ spherical harmonics.

Numerically we proceed as follows: the functions $h_n^\pm$ satisfy
\bea \label{modeequations}
\{\frac{d^2}{dr^2}+[2\kappa\frac{I_{n+1}'(\kappa r)}{I_{n+1}(\kappa r)}
+\frac{1}{r}]\frac{d}{dr}\}h_n^-(r,\kappa)
=V(r)[1+h_n^-(r,\kappa)]\kma
\\
\{\frac{d^2}{dr^2}+[2\kappa\frac{K_{n+1}'(\kappa r)}{K_{n+1}(\kappa r)}
+\frac{1}{r}]\frac{d}{dr}\}h_n^+(r,\kappa)
=V(r)[1+h_n^+(r,\kappa)]\kma
\eea
which can be solved numerically. The second differential equation is solved
starting at large $r=\bar r$ with 
$h_n^+(\bar r,\kappa)=h_n^{+'}(\bar r,\kappa)=0$, and running
backward. In principle we should take $\bar r = \infty$.
However, if$\bar r$ is chosen far outside the range of the
potential the fucntions $h_n^\pm(r,\kappa)$ are already constant
with high accuracy. In the numerical 
computation $r=\infty$ always means  $r=\bar r$ with a suitable
value of $\bar r$. 

For the first differential equation we first obtain a
solution $\tilde h_n(r,\kappa)$ 
starting at $r=0$, with $\tilde h_n(0,\kappa)=\tilde h_n'(0,\kappa)=0$.
This function does not satisfy the boundary condition 
required for the Green's function.
The function $h_n^-(r,\kappa)$ is obtained from $\tilde h_n(r,\kappa)$
in the following way: from the definition of the functions $h_n$
we have
\begin{eqnarray}\nonumber
&&f_n^-=\cali_n (1+h_n^-)\nonumber\kma\\
&&\tilde{f}_n^-=\cali_n(1+\tilde{h}_n^-)
\pkt 
\end{eqnarray}
$f_n^-$ and $\tilde f_n^-$ are both solutions of the same 
linear homogeneous differential
equation and regular at $r=0$, so they are proportional to each other,
$f_n^-=C\tilde{f}_n^-$.
The constant $C$ follows from the boundary condition at $r\to \infty$ 
as 
\begin{equation}
C=\frac{1}{1+\tilde{h}_n^-(\infty)}\kma
\end{equation}
and we obtain
\be\label{moderen}
h_n^-(r,\kappa)=\frac{\tilde h_n(r,\kappa)
-\tilde h_n(\infty,\kappa)}
{1+\tilde h_n(\infty,\kappa)}\kma
\ee
which obviously solves the differential equation with the appropriate
 boundary conditions. Of course in the numerical implementation 
$r=\infty$ is taken as $r=\bar r$ (see above).

Finally the Green's function is given by
\begin{eqnarray} \label{Green_h}
\calg( x, x',\nu^2)&=&
\frac{1}{2\pi^2}\sum_{n=0}^{\infty}(n+1)C_n^1(\cos\chi)
\cali_n(r_<,\kappa)\calk_n(r_>,\kappa)\nonumber\\
&&(1+h_n^-(r_<,\nu^2))(1+h_n^+(r_>,\nu^2))
\pkt\end{eqnarray}

In order to perform the subtractions needed in the process of
renormalization, we not only need the functions $h_n^\pm$ which are exact
to all orders in the potential $V(r)$, but also the functions
which are of first and second order, $h_n^{(1)\pm}$ 
and $h_n^{(2)\pm}$, and the inclusive sums
\be
h_n^{\overline{(m)}\pm} =\sum_{j=m}^\infty h_n^{(j)\pm}
\pkt\ee
This has been discussed in Refs. 
\cite{Baacke:1991nh,Baacke:1991sa,Baacke:1993ne}.
Obviously $h_n^\pm=h_n^{\overline{(1)}}$ as it includes 
all orders of $V(r)$ except the zero order part.
Writing the mode equations \eqn{modeequations} in the form
\be
\cald h = V(1+h)\kma
\ee
where $\cald$ denotes the differential operator on the
left hand side,we have
\bea
\cald h^{(1)}&=& V\kma\\
\cald h^{(2)}&=& Vh^{(1)}\kma\\
\cald h^{\overline{(2)}}&=&Vh^{\overline{(1)}}
\kma\eea
and so on. Solving these and similar equations we find the parts
of $h_n^\pm$ that correspond to a precise order in $V(r)$. As
$V(r)$ is of a specific order in the couplings we thus may single
out precisely specific perturbative parts.

The rescaling needed in order to pass from
$\tilde h_n^-$ to $h_n^-$ complicates affairs as it mixes orders.
After some algebra one obtains
\begin{equation}
h^{(1)-}=\tilde{h}^{(1)-}-\tilde{h}_\infty^{(1)-}
\end{equation}
and
\begin{equation}
h^{\overline{(2)}-}=\frac{\tilde{h}^{\bar{(2)}-}
-\tilde{h}_\infty^{\overline{(2)}-}
+\tilde{h}_\infty^{\overline{(1)}-}\tilde{h}_\infty^{(1)-}-\tilde{h}^{(1)-}
\tilde{h}_\infty^{\overline{(1)}-}}{1+\tilde{h}_\infty^{\overline{(1)}-}}
\pkt\end{equation}
For the Green's function we likewise may define parts of a precise
order in $V(r)$; for $ x= x'$ we have
\begin{eqnarray}
\calf( x)&=&\calg(x, x)=\frac{1}{2\pi^2}
\sum_{n=0}^\infty(n+1)^2\cali_n(r,\kappa)\calk_n(r,\kappa)\\
&&\times\left[1+h_+^{\overline{(1)}}(r,\kappa)+h_-^{\overline{(1)}}(r,\kappa)+
h_+^{\overline{(1)}}(r,\kappa)h_-^{\overline{(1)}}(r,\kappa)\right]\kma\nonumber\\
\calf^{\overline{(1)}}( x)&=&\calg^{\overline{(1)}}( x, x)
=\frac{1}{2\pi^2}\sum_{n=0}^\infty(n+1)^2\cali_n(r,\kappa)
\calk_n(r,\kappa)\\
&&\times\left[h_+^{\overline{(1)}}(r,\kappa)+h_-^{\overline{(1)}}(r,\kappa)+
h_+^{\overline{(1)}}(r,\kappa)h_-^{\overline{(1)}}(r,\kappa)\right]\nonumber
\kma\\
\calf^{\overline{(2)}}(x)&=&\calg^{\overline{(2)}}( x, x)
=\frac{1}{2\pi^2}\sum_{n=0}^\infty(n+1)^2\cali_n(r,\kappa)
\calk_n(r,\kappa)\\
&&\times\left[h_+^{\overline{(2)}}(r,\kappa)+h_-^{\overline{(2)}}(r,\kappa)+
h_+^{\overline{(1)}}(r,\kappa)h_-^{\overline{(1)}}(r,\kappa)\right]\nonumber
\pkt\end{eqnarray}
While  $\calg^{\overline{(2)}}(x, x)$ is finite, $\calg^{(0)}( x, x)$ 
and   $\calg^{(1)}(x, x)$ are divergent.
Renormalization will be discussed in section \ref{renormalization}.

%%%%%%%%%%%%%%%%%%%%%%%%%%%%%%%%%%%%%%%%%%%%%%%%%%%%%%%%%%%%%%%%%%%%%%%%%

\section{Computation of the Fluctuation Determinant}
\setcounter{equation}{0}
\label{flucdet}
The fluctuation determinant which appears in the rate formula
\be
\cald={\det}'\frac{-\partial^2+\calm^2}{-\partial^2+m^2}\kma
\ee
can be written formally as an infinite product of eigenvalues
of the fluctuation operator. The prime denotes taking the absolute
value and removing the translation mode.
As in the previous section we introduce the
generalization
\be \label{def_Dtilde}
\tilde\cald(\nu^2)=
\det\frac{-\partial^2+\calm^2+\nu^2}{-\partial^2+m^2+\nu^2}
\pkt\ee
Note that we omit the prime, here.
Using the decomposition of the Hilbert space into angular momentum
subspaces we can write
\begin{equation}
\tilde\cald(\nu^2)
=\prod_{l,n}
\left[\frac{\omega^2_{ln}+\nu^2}
{{\omega^{{}2}_{l n{(0)}}+\nu^2}} \right]
=\prod_{n=0}^\infty \left[\frac{\det \bfm_n(\nu^2) }
{\det \bfm_n^{(0)}(\nu^2)}\right]^{d_n} \kma
\end{equation}
with the radial fluctuation operators
\be
\bfm_n(\nu^2)=-\frac{d^2}{dr^2}-\frac{3}{r}\frac{d}{dr}+\frac{n(n+2)}{r^2}
+m^2+V(r)+\nu^2
\pkt
\ee
$d_n$ denotes the degeneracy, in four dimensions we have $d_n=(n+1)^2$.

According to a theorem on functional determinants of ordinary
differential operators 
\cite{Coleman85,Dashen:1974ci,Gelfand:1959nq} we can express the ratios of the
partial wave determinants via
\begin{equation}
\frac{\det \bfm_n(\nu^2)}{\det \bfm_n^{(0)}(\nu^2) } =
\lim_{r\to\infty} \frac{\psi_n(\nu^2,r)}{\psi_n^{(0)}(\nu^2,r)} \kma
\end{equation}
where $\psi_n(\nu^2,r)$ and $\psi_n^{(0)}(\nu^2,r)$ 
are solutions to equations
\begin{equation}
\bfm_n(\nu^2)\psi_n(\nu^2,r)=0 \kma~~~~~~
\bfm_n^{(0)}(\nu^2)\psi_n^{(0)}(\nu^2,r)=0 
\end{equation}
with identical regular boundary conditions at $r=0$. Of course
\begin{equation}
\psi_n^{(0)}(\nu^2,r)=\cali_n(r,\kappa) \pkt
\end{equation}
Furthermore, we have
\begin{equation}
\psi_n(\nu^2,r)=\left[1+\tilde h_n(r,\kappa)\right]
\cali_n(r,\kappa)\kma
\end{equation}
where $\tilde h_n(r,\kappa)$ has been defined in the previous section.
So we obtain
\begin{equation}\label{partialwavedet}
\frac{\det \bfm_n(\nu^2)}{\det \bfm_n^{(0)}(\nu^2) } = 
1+\tilde h_n(\infty,\kappa) \kma
\end{equation}
and
\be
\ln \tilde \cald(\nu^2)=
\sum_{n=0}^\infty d_n\ln \left[1+\tilde h_n(\infty,\kappa)\right]\pkt
\ee
The fluctuation determinant in the transition rate formula
and in the $2PPI$ formalism refers to the fluctuation
operators at $\nu^2=0$, and so in the numerical computation
we just need the functions $\tilde h_n(0,\infty)$, as for the Green's
function. The only exception is the translation mode we will
discuss in the next section.

%%%%%%%%%%%%%%%%%%%%%%%%%%%%%%%%%%%%%%%%%%%%%%%%%%%%%%%%%%%%%%%%%%%%%%%%%

\section{Unstable and translation modes}
\label{unstableandtranslation}
\setcounter{equation}{0}

In the one-loop formula for the transition rate the determinant
of the fluctuation operator appears as ${\det}' (-\partial^2+\calm^2)$,
and the prime denotes two modifications with respect to the
naive determinant: 

(i) the unstable mode has an imaginary frequency,
corresponding to a negative eigenvalue $\omega_u^2=-\nu_u^2$ of the
fluctuation operator. It is to replaced by its absolute value.
This mode appears in the $s$-wave $n=0$ and manifests itself by
a negative value of $1+\tilde h_0(0,\infty)$. So here we have to take
the absolute value when computing the fluctuation determinant.
There are no modifications of the fluctuation integral.

(ii) the translation mode manifests itself, in the semiclassical
approximation, by the asymptotic limit $1+\tilde h_1(\nu^2=0,\infty)=0$. 
When backreaction is included, there is no exact zero mode,
but a similar zero of $1+\tilde h_1(-\omega_t^2,\infty)$ persists
for a value of $\omega_t^2$ close to zero. We will identify it
as the ``would-be'' translation mode, that would be again at
$\omega_t^2=0$ in the exact theory and, accordingly, we will
treat it in analogy to the exact zero mode of the
semiclassical approximation. Of course this is justified
only as long as $\omega_i^2$ remains much smaller than
the typical energy scales such as $m^2$ or $\omega_u^2$. 

The fluctuation determinant  
\eqn{def_Dtilde} has a factor $\omega_t^2+\nu^2$
which has to be removed, according to the definition
of ${\det}'$. Otherwise the logarithm of this expression, 
appearing in the functional
determinant, does not exist.  Furthermore the Green's function
is not defined either at $\nu^2=-\omega_t^2$. 

In the semiclassical approximation the translation mode
is removed from the fluctuation determinant 
numerically in the following way \cite{Baacke:1993ne}: we
compute $\tilde h_1(\infty,\pm \epsilon^2)$ for some 
sufficiently small $\epsilon$ and replace in
Eq. \eqn{partialwavedet}
\be
\left[1+\tilde h_1(0,\infty)\right] 
 \to \frac{\tilde h_1(\epsilon^2,\infty)
-\tilde h_1(-\epsilon^2,\infty)}{2\epsilon^2}
\kma\ee
i.e., we take the numerical derivative at $\omega^2=0$.

In the backreaction computations we first determine
the position of the eigenvalue by requiring
 $1+\tilde h_1(-\omega_t^2,\infty)$ to vanish, and compute the
numerical derivative not at $\nu^2=0$ 
but at $\nu^2=-\omega_t^2$,
i.e., we remove a factor $\omega_t^2+\nu^2$.

As the Green' s function is a functional derivative
of the effective action we have to remove the zero mode
from it as well. The Green's function in the $n=1$ channel
has, at $r=r'$, the form
\be
\calg_n(r,r,\nu^2)=\frac{R_t(r)^2}{\nu^2+\omega_t^2}
+\sum_{n\neq0}
\frac{R_{1,n}^2(r)}{\nu^2+\omega_{1n}^2}
\pkt
\ee
We can use the fact that the pole term is antisymmetric 
with respect to $\nu^2+\omega_t^2$ by computing 
the Green's function
at $\nu^2=-\omega_t^2\pm \epsilon^2$ and by 
taking the average of these two values. Then the pole term has
disappeared and the averaged Green's function takes the form
\be
\frac{1}{2}\left[\calg_1(r,r,-\omega_t^2+\epsilon^2)
+\calg_1(r,r,-\omega_t^2-\epsilon^2)\right]
=\sum_{n\neq 0}R_{1n}^2(r)\frac{\omega_{1n}^2-\omega_t^2}{
(\omega_{1n}^2-\omega_t^2)^2-\epsilon^4 }\pkt
\ee
As long as $\omega_t^2$ and $\epsilon^2$ are much smaller
than the $\omega_{1n}^2$ this is a good approximation to
the desired reduced Green's function 
\be
\left[\calg_1(r,r,0)\right]_{\rm red}=\sum_{n\neq 0} \frac{R_{1n}^2(r)}{
\omega_{1n}^2}\pkt
\ee

In the explicit numerical computation of the Green's function
 we use of course the  expression \eqn{Green}.
As evident from Eqs. (\ref{moderen}) the pole 
arises from the rescaling of the mode function $\tilde h_1(\nu^2,r)$,
i.e., from dividing by $1+\tilde h_-(\nu^2,\infty)$.
In averaging over the Green's functions at $\nu^2=-\omega_t^2
\pm \epsilon^2$ we add two very large terms which almost cancel.
This can be done in a somewhat smoother way: if $\epsilon^2$
is sufficiently small we can assume that $1+\tilde h_1(\nu^2,\infty)$
passes through zero linearly and we may replace
\be
1+\tilde h_1(-\omega_t^2\pm\epsilon^2,\infty)
\to \pm\frac{1}{2}\left[\tilde h_1(-\omega_t^2+\epsilon^2,\infty)-
\tilde h_1(-\omega_t^2-\epsilon^2,\infty)\right]\pkt
\ee
The average over the Green's functions can then be cast into the form
\be
\left[\calg_1(r,r,0)\right]_{\rm red}\simeq
\frac{f_1^+(-\omega_t^2+\epsilon^2,r)\tilde f_1(-\omega_t^2+\epsilon^2,r)
-f_1^+(-\omega_t^2-\epsilon^2,r)\tilde f_1(-\omega_t^2-\epsilon^2,r)
}{\tilde h_1(-\omega_t^2+\epsilon^2,\infty)-
\tilde h_1(-\omega_t^2-\epsilon^2,\infty)}\kma
\ee
where $\tilde f_1(\nu^2,r)=I_1(\kappa r)[1+\tilde h_1(\nu^2,r)]$
is the mode function $f_1^-$ before the re-normalization.

As the translation mode is only approximate, the virial theorem
mentioned in section \ref{basicformulae} no longer holds and we have
to go back to the original expression for the prefactor, which
is the normalization of the zero mode, so we compute the 
false vacuum decay rates via
\be
\Gamma = \left(\frac{1}{2\pi\caln_0^2}\right)^2
\exp(-S_{\rm eff})
\ee
Of course the treatment of the approximate zero mode is an additional
approximation, beyond the one-loop or Hartree approximations. 
In the numerical computations the squared frequency of the zero mode
$\omega_t^2$ is generally of the order $10^{-3}$ in mass units. 
It becomes at most of order
$10^{-1}$ near the critical points where our iteration ceases
to converge, see section \ref{numericalresults}. 
So the approximation seems to be justified.

%%%%%%%%%%%%%%%%%%%%%%%%%%%%%%%%%%%%%%%%%%%%%%%%%%%%%%%%%%%%%%%%%%%%%%%%%

\section{Renormalization}
\label{renormalization}
\setcounter{equation}{0}

In the previous sections we have presented the basic formalism and
its numerical implementation. There is still one point to be discussed:
divergences and renormalization.
As we are able to single out precise perturbative orders of the
Green's functions and of the functional determinants, we
can do the regularization of the leading orders
analytically; we will use dimensional regularization.
In particular we have
\be \label{greenpert}
\calf( x) =\calf^{(0)}(x)+
\calf^{(1)}(x)+
\calf^{\overline{(2)}}( x)
\ee
and
\be\label{detpert}
\cald=\cald^{(1)}+\cald^{(2)}+\cald^{\overline{(3)}}
\kma\ee
where in both equations the first and second terms are divergent
and the last term is convergent. The convergent parts are computed
numerically using the methods described in the previous sections.
The divergent terms are first computed and renormalized analytically.
Then their finite parts are evaluated numerically, this is straightforward
and involves some Fourier-Bessel transforms of the classsical
profiles.

Renormalization not only needs regularization but also renormalization
conditions. In some previous publications 
\cite{Baacke:2003uw,Baacke:2004xk} we have used $\overline{MS}$
subtraction which combines regularization and renormalization.
Here we will impose renormalization conditions
which keep the effective potential close to
the tree level one. For comparison we will also present 
results obtained in the $\overline{MS}$ scheme.

The effective potential does not play any
r\^{o}le in our computations, we always use the {\em effective action}.
The effective potential is obtained if the classical field is homogeneous
in space and time, or, as relevant here, in $4$-dimensional Euclidean
space. The effective potential is often used when discussing 
quantum corrections, e.g., to Higgs potentials. It is relevant for the
properties of the vacua, for double well structures it becomes
complex around the maximum of the potential between the two wells.

We now discuss the renormalization conditions.
The false vacuum is relevant for the asymptotic region of the bounce
as $r\to\infty$. It is convenient to require that in this region
$\phi\to 0$. This implies that the left minimum of the effective potential,
for one-loop or Hartree back-reactions, remains at $\phi=0$, and
that the mass remains the bare mass:
\bea \nonumber
U_{\rm eff}(0)&=&U(0)=0\kma
\\\label{renormcond1}
U'_{\rm eff}(0)&=&U'(0)=0\kma
\\\nonumber
U''_{\rm eff}(0)&=& m^2\pkt
\eea
It is reasonable, furthermore, to require that the vacuum expectation
value in the true vacuum  retains its tree level value, and that
the energy difference $\epsilon=-U(\phi_{\rm tv})$ 
between the two vacua retains its tree level value,
so that the semiclassical, one-loop and Hartree aproximations refer
to essentially the same physical situation. Of course the exact
shape of the effective potentials is different for these three 
cases. So we require
\bea\nonumber
U'_{\rm eff}(\phi_{\rm tv})&=&0\kma\\
U_{\rm eff}(\phi_{\rm tv})&=&U(\phi_{\rm tv})=-\epsilon\pkt
\label{renormcond2}\eea
The renormalization of the effective potential is discussed in
Appendix \ref{appendix:effpot}. For the one-loop back-reaction
the counterterm potential is given by
\be
\delta U=-\delta \call= \delta\rho \phi+
\frac{1}{2}\delta m^2 \phi^2-\delta \eta \phi^3
+\frac{\delta \lambda}{8}\phi^4
\pkt\ee
Using the definition
\be
L_\epsilon= \frac{2}{\epsilon}-\gamma_E+\ln 4\pi-\ln\frac{m^2}{\mu^2}
\ee
we find
\bea\label{delrho1loop}
\delta \rho &=& -\frac{3\eta m^2}{16\pi^2}(L_\epsilon+1)\kma 
\\
\delta m^2 &=&\frac{3}{32\pi^2}(12\eta^2+\lambda m^2)L_\epsilon
+\frac{3\lambda m^2}{32\pi^2}\kma
\\
\delta \eta &=&18\eta\lambda\frac{L_\epsilon}{64\pi^2}+
\delta\eta_{\rm fin}\kma\\
\delta \lambda &=&\frac{9\lambda^2}{32\pi^2}L_\epsilon
+\delta\lambda_{\rm fin}
\pkt\eea
Here the finite terms $\delta\eta_{\rm fin}$ and $\delta\lambda_{\rm fin}$
are the solutions of a linear system of equations given in 
Appendix \ref{appendix:effpot}. With these counterterms the equations
of motion and the effective action of the bounce become finite.
The relevant equations are given in Appendices
\ref{renormequmot1loop} and \ref{renormaction1loop}. In the $\overline{MS}$
scheme the counter terms only consist of the parts proportional 
to $L_\epsilon$ and all the remaining finite parts  are 
set to zero. However, we have to deviate slightly from these conventions, 
$\delta \rho$ has to be chosen as in Eq. \eqn{delrho1loop}, otherwise
the false vacuum is shifted away from $\phi=0$. The other finite parts 
have been set to zero.

For the renormalization of the  Hartree back-reaction it is essential
that for a mass-independent regularization scheme 
all divergent parts are related  \cite{Verschelde:2000dz,Verschelde:2000ta}.
As a consequence these divergences
 can be conveniently removed by one counter term \cite{Nemoto:1999qf}
\be
\delta U_{\rm div}= B(\calm^4-m^4)=
\frac{1}{64\pi^2} (L_\epsilon+1)  (\calm^4-m^4)
\kma\ee
see also the discussion in the Appendix of \cite{Baacke:2003bt}. The part
proportional to $m^2$ is an infinite renormalization of the vacuum
energy. For the $\overline{MS}$ scheme this is all we have to do
\footnote
{In the strict sense $B$ should  be chosen equal to $L_\epsilon/16\pi^2$ 
in the $\overline{MS}$ scheme. This would lead to some tedious modifications
of the back-reaction calculations. However, one can always change 
$L_\epsilon$ to $L_\epsilon+1$ by modifying the renormalization scale $\mu$.}. 
If we want to impose
the same boundary conditions as in the one-loop approximation 
with back-reaction, we have to introduce a set of finite renormalizations
\be
\delta U_{\rm fin}=\delta \Lambda_{\rm fin}+\delta \rho_{\rm fin}\phi
+\frac{1}{2}\delta m^2_{\rm fin}\phi^2-
\delta \eta_{\rm fin}\phi^3+\frac{1}{8}\delta \lambda_{\rm fin}\phi^4
\pkt\ee
Imposing again the conditions \eqn{renormcond1} and \eqn{renormcond2}
we find
\be
\delta \Lambda_{\rm fin}=\delta\rho_{\rm fin}=\delta m^2=0
\pkt\ee
Fixing the remaining counter terms becomes more involved; due to the
nonlinearity of the gap equation we get a
set of nonlinear equations. An iterative procedure is used to fix 
$\delta \eta$ and $\delta \lambda$ numerically. This is
discussed in Appendix \ref{renormeffpothartree}.
For the $\overline{MS}$ scheme the finite counter terms are
set to  zero, $\delta U_{\rm fin}\equiv 0$. 

With these counter terms the dynamical equations of the Hartree
scheme become finite. The finite equations are given in detail
in Appendix \ref{renormhartree}.

%%%%%%%%%%%%%%%%%%%%%%%%%%%%%%%%%%%%%%%%%%%%%%%%%%%%%%%%%%%%%%%%%%%%%%%%%

\section{Numerical results}
\setcounter{equation}{0}
\label{numericalresults}
\subsection{General remarks}

In discussing the numerical results it is convenient to use a
parametrization which weights the relative importance of the
classical and quantum parts of the action. One introduces 
the parameters
\bea
\beta&=&\frac{m^4}{4\eta^2}\kma\\
\alpha&=&\lambda\beta
\kma\eea
and the rescaling of the fields $\phi=\sqrt{\beta}\hat \phi$.
As to the third parameter in the potential, the mass $m$,
we have set it equal to unity in our numerical computations.
So all dimensionful parameters, like $\eta$ or $\epsilon$,
and all dimensionful results are understood to 
be given in mass units.

The parameter $\alpha$ parametrizes the shape of the potential,
for small $\alpha$ the potential is strongly asymmetric, 
for $\alpha=1$ it becomes a symmetric double well potential.
For values of $\alpha$ larger than $1$ the r\^ole of true and false
vacua is interchanged, so we can restrict ourselves to
$0<\alpha<1$.

In the semiclassical approximation, i.e., one-loop without
back-reaction one has
\be
S_{\rm eff}=\beta \hat S_{\rm cl}(\hat \phi)+ S_{\rm 1-loop}
\kma\ee
where $\hat S_{\rm cl}(\hat \phi)$ and $S_{\rm 1-loop}$ only depend
on $\alpha$; so for large $\beta$ the action is dominated
by the classical part and for small $\beta$ by the quantum part.
Of course this is not strictly true once one introduces back-reaction.
Still we expect the effects of back-reaction to be small
for large $\beta$ and important for small $\beta$. Furthermore,
for large $\beta$ the tunneling rate gets strongly suppressed.
So our main interest will be in small values of $\beta$.

The self-consistent profiles $\phi(r)$ are obtained by iteration, each step
of iteration consists in solving for a given
$\calf(r)$ the equation for the profile $\phi(r)$, and by computing
then the new values of $\calf(x)$ for this profile.
Of course in the first step $\calf(r)$ is not yet known; if
the quantum corrections are large it is  precarious
to start the first iteration with $\calf(r)\equiv 0$. 
In order to avoid this problem
we have computed, at fixed $\alpha$, a series of solutions starting
with a large value of $\beta$, where the corrections are small. 
Then with descending
values of $\beta$ we have used in the first iteration step the
self-consistent values of $\calf(r)$ of the previous
value of $\beta$. 

For the discussion of the data it is useful to introduce a shorthand
notation for the four different cases we have investigated:
\begin{itemize}
\item[-] Case $I$: one-loop back-reaction, renormalization
conditions Eqs. \eqn{renormcond1} and \eqn{renormcond2}, i.e.
preserving the parameters  of the true vacuum.
\item[-] Case $II$: one-loop back-reaction, $\overline{MS}$ 
renormalization.
\item[-] Case $III$: Hartree back-reaction, renormalization
conditions Eqs. \eqn{renormcond1} and \eqn{renormcond2}
\item[-] Case $IV$: Hartree back-reaction, $\overline{MS}$ 
renormalization.
\end{itemize}

\begin{figure}[htbp]
  \centering
\vspace{7mm}
   \includegraphics[scale=1]{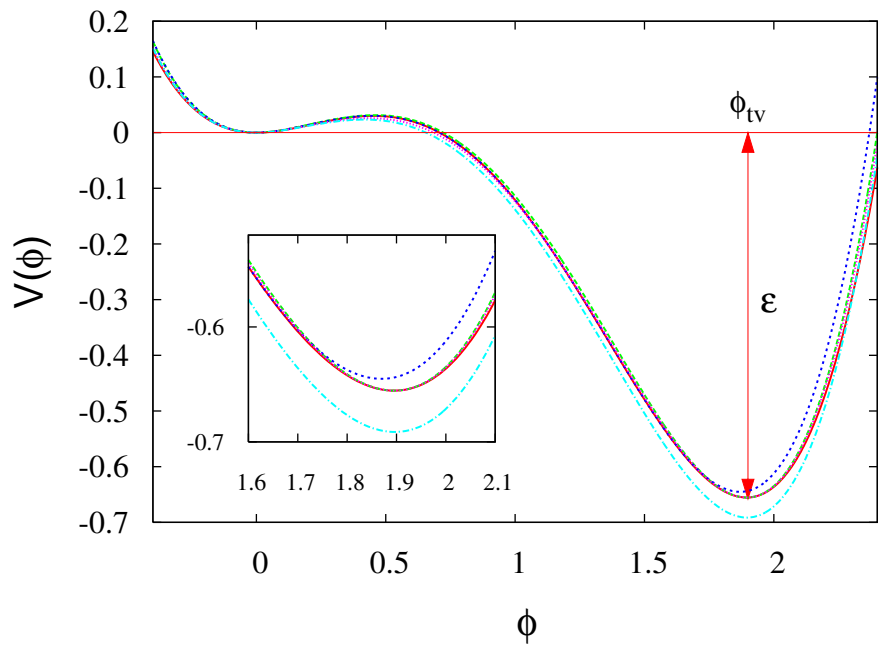}
\vspace{4mm}
  \caption{The effective potential for $\alpha=0.7$ and $\beta=0.3$.
solid line: tree level 
potential; dotted line: case $I$; dash-dotted line: case $II$; short-dashed 
line: case $III$; long-dashed line: case $IV$; cases $I$ and $III$ are
hardly visible.}
  \label{figure:potential}
\end{figure}

We display, in Fig. \ref{figure:potential}
the effective potential for $\alpha=0.7$ and $\beta =0.3$ for the
cases $I$ to $IV$. The effective potential for cases $I$ and $III$
is very close to the tree level potential over the whole range
presented in the figure. The position of the true vacuum is seen
to be shifted in different ways in the one-loop and Hartree cases.

%------------------------------------------------------------------------

\subsection{The bounce profiles}
\begin{figure}[htbp]
  \centering
\vspace{7mm}
   \includegraphics[scale=.7]{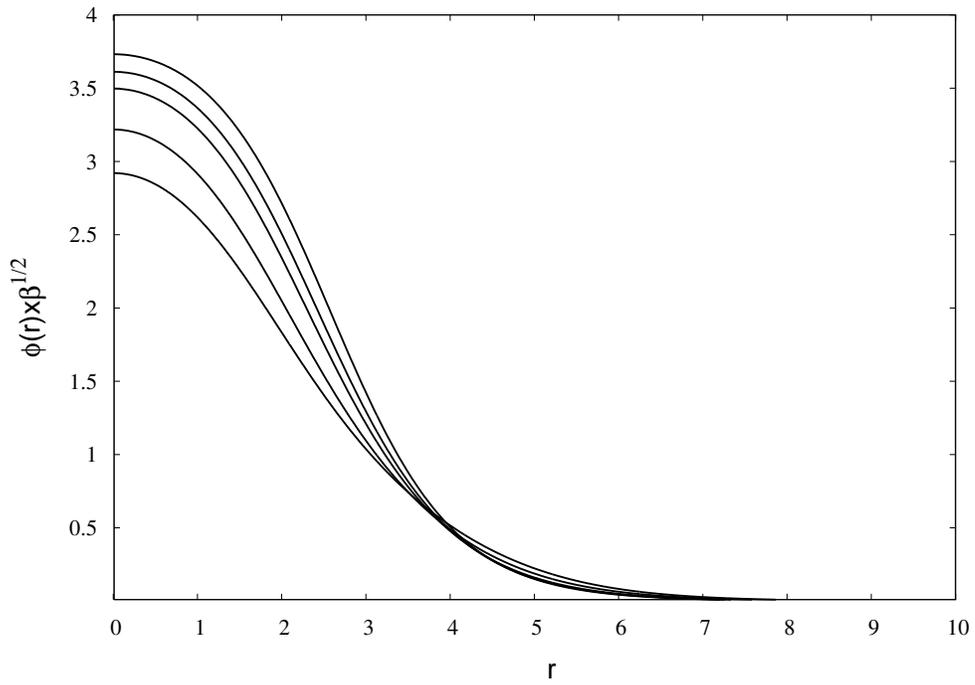}
\vspace{4mm}
  \caption{Behavior of $\hat \phi(r)=\phi(r)/\sqrt{\beta}$ 
for  $\beta = 20, 1 , 0.5, 0.25$ and $0.2$ at $\alpha=0.6$,
for case III. In the central region,
$r \lesssim 4$ these profiles are seen to decrease with 
decreasing $\beta$ while
they are independent of $\beta$ in the semiclassical approximation. }
  \label{figure:normalizedprofiles}
\end{figure}

In the semiclassical approximation the profiles $\phi(r)$ are, at fixed
parameter $\alpha$ and arbitrary values of $\beta$, 
determined by one universal profile
\be
\hat \phi_\alpha(r)=\beta^{-1/2}\phi(r)
\ee
which is independent of $\beta$.
Therefore, the change of the profile by the back-reaction can
be displayed in a transparent way by plotting,
at fixed $\alpha$, the corresponding
normalized profiles
\be
\hat \phi(r)=\beta^{-1/2}\phi(r)
\ee
for different values of the parameter $\beta$. For large $\beta$, when the 
backreaction is weak, these profiles are expected to be independent
of $\beta$ and close to
$\hat\phi_\alpha(r)$. Indeed this is what we observe
for $\beta \gtrsim 10$. The $\beta$-dependence observed for
smaller values of $\beta$ depends on the type of back-reaction and on
the renormalization conditions. For the presentation
in Fig. \ref{figure:normalizedprofiles}  we have chosen
 the Hartree approximation and
the renormalization conditions which preserve the position of the
true vacuum (case $III$). With decreasing $\beta$ the normalized
profiles get lower and lower in the central
region of the bounce. If $\beta$ becomes smaller than $1$, this
decrease becomes substantial. For some lower limiting value 
of $\beta \simeq 0.2$ the iterative procedure ceases to converge, 
during the iteration the profile collapses to $\phi(r)\equiv 0$.
The qualitative behavior is similar for all other cases and all parameters
$\alpha$ we have considered ($0.2, 0.3, \dots 0.8$).

\begin{figure}[htbp]
  \centering
\vspace{7mm}
   \includegraphics[scale=.7]{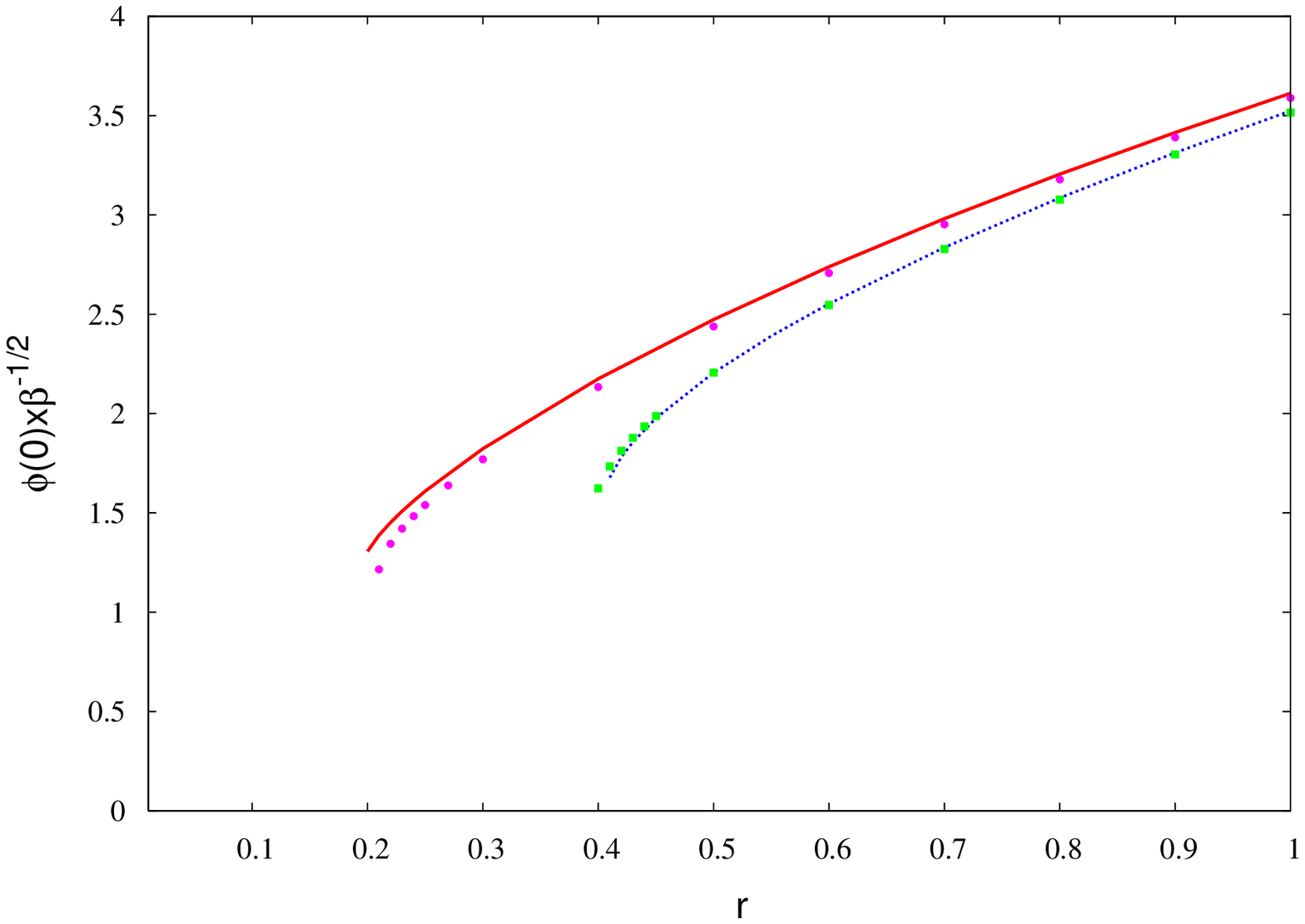}
\vspace{4mm}
  \caption{Behavior of $\phi(0)/\sqrt{\beta}$ near the critical
values which limit the region of convergence for
$\alpha=0.6$. Short-dashed line: case $I$; 
full squares: case $II$; solid line: case $III$ and
full circles: case $IV$.}
  \label{figure:phi0abs}
\end{figure}

In Fig. \ref{figure:phi0abs} we show the behavior of $\phi(0)/\sqrt{\beta}$
near the critical values of $\beta$ for the four different
cases. This ratio becomes constant as $\beta \to \infty$ and would be
constant throughout in the semiclassical approximation.
The behavior near the critical values is seen to be singular, as far
as one can tell from a numerical computation. We therefore
think that this phenomenon is genuine and not related to the way in which
we iterate the equations. We have tried to modify the iteration by using
the standard ``overrelaxation'' scheme
\be
f_{i+1}(r)=\sigma f_i(r)+ (1-\sigma)(\calo f_i)(r)
\ee 
where $f_i(r)$ is the $i$'th iteration of $\phi(r)$ and
$\calm(r)$, and $\calo$ is the operation that generates the new
profiles in the normal iterative process ($\sigma=0$). 
The runaway of the iteration below the critical value
of $\beta$ was persistent for various parameters $\sigma$ that we have tried
out. The tendency towards $\phi\equiv 0$ 
is apparent already for the parameter values of $\beta$ where the iteration
still converges, see Fig. \ref{figure:normalizedprofiles}.

For $\alpha \lesssim 0.6$ and Hartree back-reaction
the critical values of $\beta$ are around $\beta \simeq 0.2$ 
and then increase strongly. For the one-loop
backreaction they increase roughly linearly from $0.2$ to $0.6$ 
between $\alpha=0.3$ and $\alpha=0.7$. At $\alpha=0.8$ the critical
value is around $\beta=6$ (!) for all cases.

%------------------------------------------------------------------------

\subsection{Effective actions and transition rates}

As mentioned above, the effective action in the semiclassical case 
is of the form
\be
S_{\rm eff}=\beta \hat S_{\rm cl}[\hat \phi] + S_{ 1-\rm loop}
\kma\ee
where $\hat S_{\rm cl}$ and $S_{ 1-\rm loop}$ are functions of
$\alpha$ only. So we may display the effect of the back-reaction
by plotting
\be
\rho=\frac{S_{\rm eff}}{\beta \hat S_{\rm cl}[\hat \phi] 
+ S_{ 1-\rm loop}}
\kma\ee 
where $S_{\rm eff}$ is the effective action in the various
approximations and renormalization conventions. $S_{ 1-\rm loop}$
in the denominators depends on the renormalization and is therefore
different in the various cases. Clearly the ratio $\rho$ should go to
unity as $\beta \to \infty$, because there $\eta=1/2\sqrt{\beta}$
and $\lambda=\alpha/\beta$ go to zero. Actually already for $\beta \gtrsim 1$
$\rho$ is very close to one. Even near the point where the iteration
ceases to converge, the deviation from unity is only a few percent.

\begin{figure}[htbp]
  \centering
\vspace{7mm}
   \includegraphics[scale=.7]{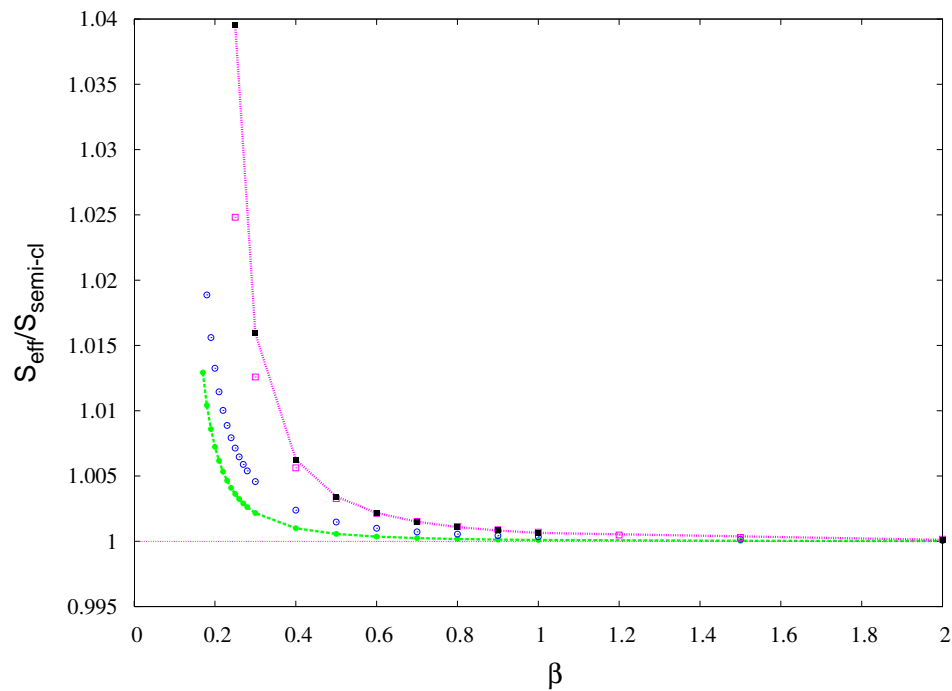}
\vspace{4mm}
  \caption{Ratio of $S_{\rm eff}$ and $S_{\rm semi-cl}$ as a function
of $\beta$ for $\alpha=0.3$; dotted line with full squares: case I;
empty squares: case II; long-dashed line with full circles: case III; 
empty circles: case IV.  }
  \label{figure:rho.3}
\end{figure}

\begin{figure}[htbp]
  \centering
\vspace{7mm}
   \includegraphics[scale=.7]{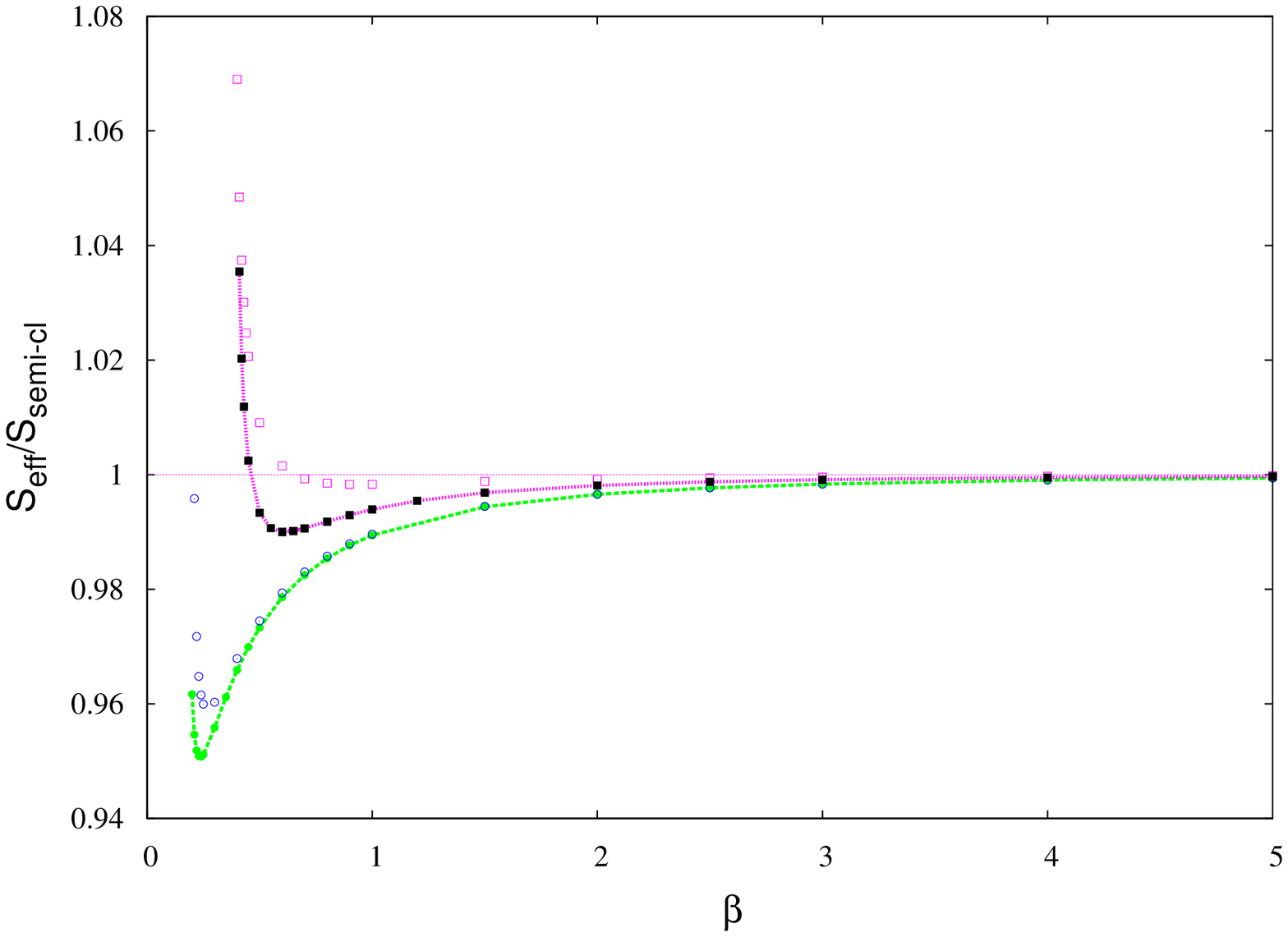}
\vspace{4mm}
  \caption{Ratio of $S_{\rm eff}$ and $S_{\rm semi-cl}$ as a function
of $\beta$ for $\alpha=0.6$; dotted line with full squares: case I;
empty squares: case II; long-dashed line with full circles: case III; 
empty circles: case IV.}  \label{figure:rho.6}
\end{figure}

We display $\rho$ in Figs. \ref{figure:rho.3} and Figs.
\ref{figure:rho.6}. Both figures show the deviations for the four different
cases.  We see that the deviations for $\beta \gtrsim 1 $ are quite
small; for $\beta=0.3$ $\rho$ is larger than $1$ for all $4$ cases,
for $\beta=0.6$ we have $\rho < 1$ for $\beta > 0.5 $, for smaller 
values it increases strongly and displays some singlarity, a
cusp or pole, near the critical value.
The semiclassical action is dominated by the classical
action, and so is the effective action including the various backreactions.
We have, e.g., for $\alpha=0.6$ and case II, 
$S_{\rm semi-cl}=395.08\beta-29.522$ where the first term is the 
classical action, and the second one the one-loop action.
So the contribution of the quantum action is relatively small, and even
substantial changes would not affect $\rho$. Indeed near the critical
point it is the classical action which deviates strongly from its
semiclassical value, and this is due to the strong changes in the
profile $\phi(r)$. We have to stress that here we are considering
{\em relative} changes of the effective action. As the effective
actions are of the order of a few hundred this implies absolute
changes of several units, and this is what enters the transition
rates.

Another quantity of interest is, therefore, the ratio between 
the transition rates
and their semiclassical values. In part this reflects the behavior
of the effective actions, but also includes the ratio of
the prefactors. For large $\beta$ these ratios will again go to unity,
the logarithm will  go to zero. 
We plot the logarithm of $\Gamma/\Gamma_{\rm semi-cl}$ for 
small values of $\beta$, for $\alpha=0.3$ and $\alpha=0.6$.
We see that the ratio can amount to several orders in magnitude.
for  the one-loop backreaction. the rates are always suppressed
with respect to the semiclassical rate. This is what we found as well
in the $2$-dimensional case \cite{Baacke:2004xk} for the Hartree
backreaction. Here we find, for the Hartree case, suppression for 
$\beta \lesssim 0.5$ and enhancement for larger values of $\alpha$.

\begin{figure}[htbp]
  \centering
\vspace{7mm}
   \includegraphics[scale=.7]{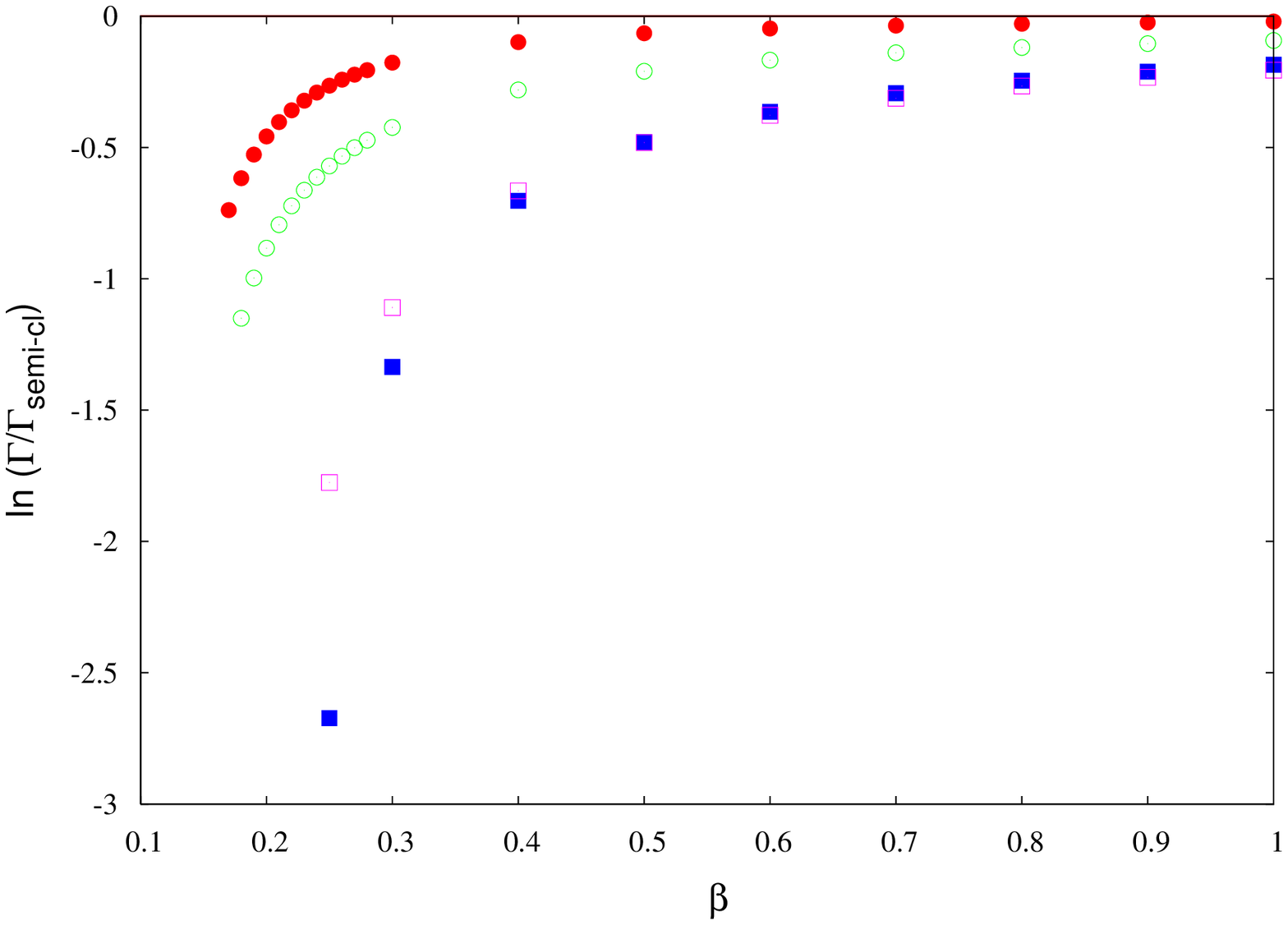}
\vspace{4mm}
  \caption{Logarithm of the ratio $\Gamma/\Gamma_{\rm semi-cl}$
as a function
of $\beta$ for $\alpha=0.3$; full squares: case I;
empty squares: case II; full circles: case III; 
empty circles: case IV.}  \label{figure:gamma.3}
\end{figure}

\begin{figure}[htbp]
  \centering
\vspace{7mm}
   \includegraphics[scale=.7]{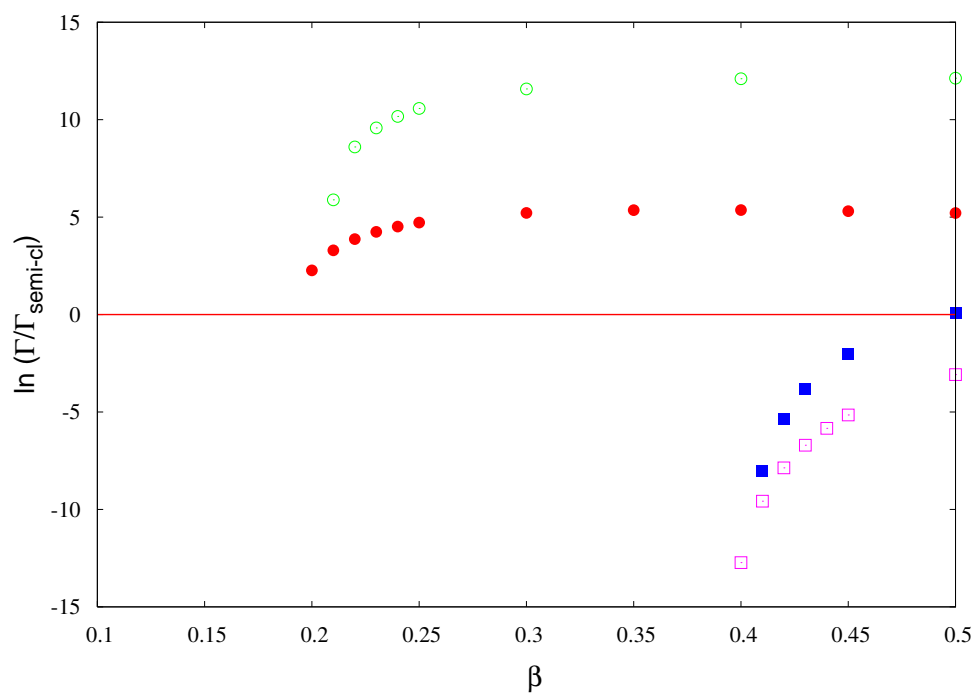}
\vspace{4mm}
  \caption{Logarithm of the ratio $\Gamma/\Gamma_{\rm semi-cl}$
as a function of $\beta$ for $\alpha=0.6$;
full squares: case I;
empty squares: case II; full circles: case III; 
empty circles: case IV.}  \label{figure:gamma.6}
\end{figure}

We have not been able to pinpoint precisely the reason for 
disappearance of the bounce solution at small $\beta$. We can add the following
comments:
\begin{itemize}
\item[-] the fluctuation integral shows no anomaly in the critical region.
This means in particular that the instability of the
classical solution is not caused by the appearance of a
further unstable
mode in the fluctuation operator. If the squared frequency of such a
mode would cross zero, the fluctuation integral would receive a very
large contribution proportional  to the square of its
wave function. We display the fluctuation integral $\alpha=0.5$ and
for various values of $\beta$ in Fig. \ref{figure:fluctuationintegral}.
In the semiclassical approximation it is independent of $\beta$,
here we clearly see a $\beta$-dependence which becomes stronger
for small $\beta$, but there is no sign of a singularity in this behaviour
near the critical value $\beta\simeq 0.2$. In the equation of motion 
for $\phi$ this integral is multiplied by $\lambda\phi=\alpha\phi/\beta$, so
this contribution becomes more and more important for small $\beta$.

\begin{figure}[htbp]
  \centering
\vspace{7mm}
   \includegraphics[scale=.7]{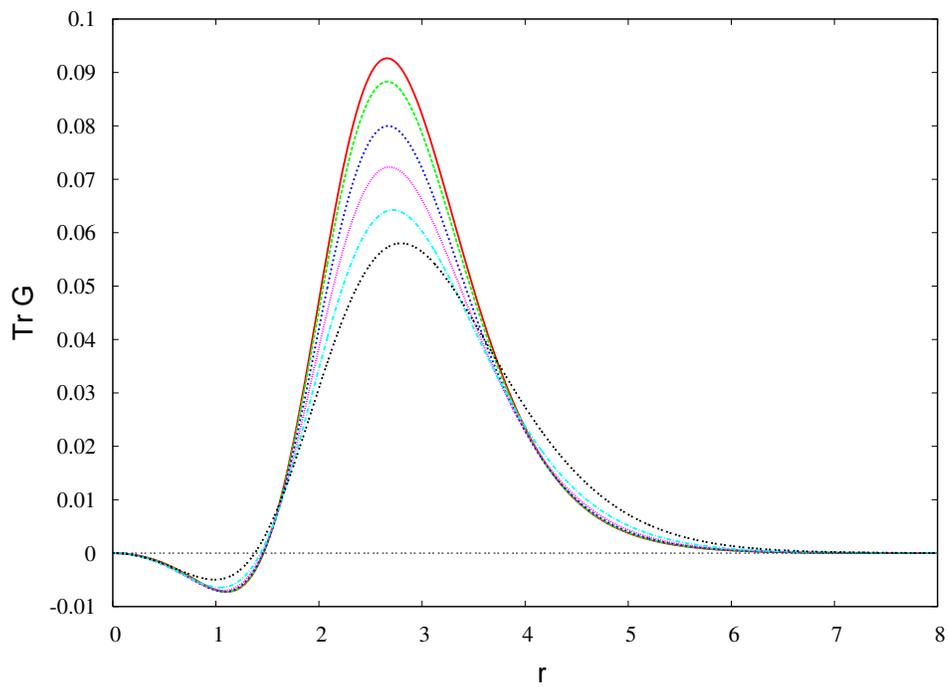}
\vspace{4mm}
  \caption{The fluctuation integral $\calf(r)$ for the Hartree
backreaction, case $III$, as a function of $r$ 
for $\alpha=0.5$. The curves display a decrease with $\beta$, which takes
the values $10, 2.5, 0.9, 0.5, 0.3$ and $0.2$.} 
\label{figure:fluctuationintegral}
\end{figure}

\item[-] the deviations of $S_{\rm eff}$ from the semiclassical action
are mostly due to the changes in the classical action, induced by
the change of the profiles $\phi(r)$. The one-loop and double-bubble
contributions to the effective action do change substantially, but
their influence on $S_{\rm eff}$ remains numerically small.
\item[-] the convergence of the partial wave summation both for
the fluctuation determinant and for the fluctuation integral is very good;
we have included terms up to $n=20$  and included the asymptotic
tail by an extrapolation (see Refs. \cite{Baacke:2003uw,Baacke:2004xk}).
The convergence remains excellent  near the critical values of $\beta$.
\item[-] we have not spent too much effort in studying the
region $\alpha > 0.7$, where we approach the thin-wall regime.
The classical actions become so large there that the semiclassical
action is of the order of a few thousands (e.g. $1500\beta$ for
$ \alpha=0.8$) and correspondingly the transition rates are
incredibly small. We note, nevertheless, that for $\alpha =0.8$
the critical values of $\beta$ are around $6$ in the various
cases we have studied.  
\end{itemize}
We think that the instability of the bounce solution is mainly due to
the term $\lambda\phi(r)\calf(r)$ in the differential equation for the
bounce. Its effects are difficult to analyze: we do not simply have
a change of the effective potential, as displayed in
Fig. \ref{figure:potential}, but a modification that depends on $r$.

%%%%%%%%%%%%%%%%%%%%%%%%%%%%%%%%%%%%%%%%%%%%%%%%%%%%%%%%%%%%%%%%%%%%%%%%%

\section{Summary and conclusions}
\setcounter{equation}{0}
\label{summary}

We have presented here two schemes for incorporating quantum backreaction
into the computation of transition rates in quantum field theory, here
applied to false vacuum decay via a bounce: 
a one-loop backreaction, where
the bounce profile is computed so as to minimize the effective
action instead of the classical action; and the Hartree
backreaction where the quantum backreaction is included into
the computation of the profile {\em and} of the quantum
fluctuations. We have derived the general equations including their
finite, renormalized form. We have used dimensional regularization
and applied two different sets of renormalization conditions.

We have presented numerical results for different parameter sets,
using the $\alpha-\beta$ parametrization of Ref. \cite{Dine:1992vs}. 
The corrections to the semiclassical transition rate remain small
as long as $\beta=1/4\eta^2$ is greater than $0.6$ and $\alpha=\lambda\beta$
is smaller than $0.7$. For lower values of $\beta$ the deviations
from the semiclassical transition rate become sizeable and can
amount to several orders of magnitude. These deviations depend on the
approximation, one-loop or Hartree, and on the renormalization conditions.
The transition rates are reduced for the one-loop backreaction, for 
the Hartree backreaction they are reduced for $\alpha \lesssim 0.5$ and
enhanced for larger values of $\alpha$.

 We find  a critical value
of $\beta$, below which our iteration ceases to converge. The behavior of
various quantities near this value suggests that this is a genuine
phenomenon and not a technical deficiency of our iteration scheme. If 
this is so, it implies that bounce solutions do not exist below this
critical value and that in this region the false vacuum decay 
proceeds via different configurations in functional space.
This is not unexpected for strong couplings:
$\eta \gtrsim 1$ and $\lambda \gtrsim \alpha/0.3$.

Clearly, the computation of transition rates using the semiclassical
approximation and with quantum backreaction remain approximations which
cannot be compared, at present, with experimental results or
with lattice computations, such as
those in Refs. \cite{Alford:1993zf,Alford:1993ph,Moore:2001vf}.
 So it is not clear to what extent the inclusion
of one-loop corrections, with or without backreaction, really
represents an improvement. We may infer from our results that the
semiclassical approximation is stable with respect to these
higher corrections for $\alpha \lesssim 0.7$ and $\beta \gtrsim 0.6$, so
we may believe it to be reliable. For lower values
of $\beta$ the deviations are strong, especially in the transition rates
they amount to a few orders of magnitude. As the different approximations
lead to quite different results, a comparison with lattice 
data would be of great interest.

The methods presented here can be applied to various other 
computations of quantum backreaction, e.g. to the 
selfconsistent computation of classical solutions which minimize
the sum of classical and zero point energies. While there 
one does not use the determinant theorem, one likewise employs techniques
based on Green's functions in Euclidean space  
\cite{Baacke:1991nh,Baacke:1998nm,Baacke:2000bv}
or techniques based on the analysis of phase shifts 
\cite{Graham:2002fw,Graham:2002xq,Graham:2002fi}.
Other possible applications of such selfconsistent computations
include quantum corrections 
to instantons \cite{Baacke:1994bk,Burnier:2005he}, vortices
\cite{Goodband:1995qs,Nagasawa:2002at,Bordag:2002sa} and other
classical solutions in quantum field theory. There are even cases
without such a classical solution as, e.g., the selfconsistent
pion cloud in the chiral quark model  
\cite{Diakonov:1987ty,Christov:1995vm,Alkofer:1994ph,Baacke:1998nm}.

\section*{Acknowledgments}

N. K.  thanks the 
\emph{Deutsche Forschungsgemeinschaft} 
for financial support as a member 
of \emph{Gra\-du\-ier\-ten\-kol\-leg 841}.

%%%%%%%%%%%%%%%%%%%%%%%%%%%%%%%%%%%%%%%%%%%%%%%%%%%%%%%%%%%%%%%%%%%%%%%%%
%%%%%%%%%%%%%%%%%%%%%%%%%%%%%%%%%%%%%%%%%%%%%%%%%%%%%%%%%%%%%%%%%%%%%%%%%

\begin{appendix}

\section{Renormalization of the effective potential}
\label{appendix:effpot}
\setcounter{equation}{0}

\subsection{The one-loop effective potential}
In the one-loop approximation the Green's function are computed
with the effective mass
 \begin{equation}
\calm^2=U''(\phi)=m^2-6\eta\phi+\frac{3}{2}\lambda\phi^2
\pkt\end{equation}
Including the classical potential and the counterterms
the 1-loop effective potential is given by
\begin{eqnarray}
U_{\rm eff}(\phi)&=&U(\phi)
+\frac{1}{2}\ln\det\frac{-\partial^2+\calm^2}{-\partial^2+m^2}
+\delta U(\phi)\nonumber\\
&=&\frac{m^2}{2}\phi^2-\eta\phi^3+\frac{\lambda}{8}\phi^4
-\frac{\calm^4}{64\pi^2}\left(L_\epsilon
-\ln\frac{\calm^2}{m^2}+\frac{3}{2}\right)
+\frac{m^4}{64\pi^2}\left(L_\epsilon+\frac{3}{2}\right)\nonumber\\
&&+\delta\rho \phi+\frac{1}{2}\delta m^2\phi^2-\delta\eta\phi^3
+\frac{1}{8}\delta\lambda\phi^4\nonumber\\
\pkt\end{eqnarray}
For the first and the second derivative of effective potential
one finds 
\begin{eqnarray}
U_{\rm eff}'(\phi)&=&m^2\phi-3\eta\phi^2+\frac{\lambda}{2}\phi^3+\delta\rho
+\delta m^2\phi-3\delta\eta\phi^2
\\
&-&\frac{\calm^2}{32\pi^2}(3\lambda\phi-6\eta)
\left(L_\epsilon-\ln\frac{\calm^2}{m^2}+1\right)\kma\nonumber\\
U_{\rm eff}''(\phi)&=&m^2-6\eta\phi+\frac{3}{2}\lambda\phi^2+\delta m^2-6\delta \eta
\phi+\frac{3}{2}\delta\lambda\phi^2\\ \nonumber
&-&\frac{1}{32\pi^2}(3\lambda\phi-6\eta)^2
\left(L_\epsilon -\ln\frac{\calm^2}{m^2}\right)
-\frac{3\lambda\calm^2}{32\pi^2}\left(L_\epsilon -\ln\frac{\calm^2}{m^2}+1\right)\pkt
\end{eqnarray}
We have already $U_{\rm eff}(0)=0$. The vanishing of the first derivative
at $\phi=0$ fixes
\be
\delta \rho=-\frac{12\eta m^2}{64\pi^2}(L_\epsilon+1)\pkt
\ee
The condition $U''(0)=m^2$ leads to
\be
\delta m^2=\frac{3}{32\pi^2}(12\eta^2+\lambda m^2)L_\epsilon
+\frac{3\lambda m^2}{32\pi^2}
\pkt\ee
The absolute minimum (true vacuum) of the classical potential occurs at
\begin{equation}
\phi_{\rm tv}=\frac{3\eta}{\lambda}+\sqrt{\frac{9\eta^2}{\lambda^2}-\frac{2m^2}{\lambda}}
\pkt\end{equation}
If its position (the vacuum expectation value)
is to be retained we have to  require $U'_{\rm eff}(\phi_{\rm tv})=0$.
If the energy difference $\epsilon$ between the vacua is put equal to its
tree level value we have to impose
\be
U_{\rm eff}(\phi_{\rm tv})=U(\phi_{\rm tv})
\pkt\ee
The latter two conditions lead to 
\begin{eqnarray}
\delta \eta&=&9\eta\lambda\frac{L_\epsilon}{32\pi^2}+\delta\eta_{\rm fin}\kma\\
\delta\lambda&=&\frac{9\lambda^2}{32 \pi^2}L_\epsilon+
\delta \lambda_{\rm fin}\kma
\end{eqnarray}
where the finite parts satisfy the
linear system of equations
\begin{eqnarray} \label{etalambdaeqs1loop}
&&\hspace{-8mm}-\delta \eta_{\rm fin}\phi_{\rm tv}^3+\frac{1}{8}\delta\lambda_{\rm fin}\phi_{\rm tv}^4
\\\nonumber
&&=\frac{1}{64\pi^2}
\left[-\calm_{\rm tv}^4\left(\ln\frac{\calm_{\rm tv}^2}{m^2}-\frac{3}{2}\right)
-3\lambda m^2\phi^2_{\rm tv}+12 \eta m^2\phi_{\rm tv}
-\frac{3}{2}m^4\right]\kma\nonumber\\
&&\hspace{-8mm}-3\delta\eta_{\rm fin}m^2\phi_{\rm tv}^2+\frac{1}{2}\delta\lambda_{\rm fin}\phi_{\rm tv}^3 
\\\nonumber
&&=\frac{6\lambda\phi_{\rm tv}-12\eta}{64\pi^2}\left[-\calm_{\rm tv}^2
(\ln\frac{\calm_{\rm tv}^2}{m^2}-1)-m^2\right]\nonumber\kma
\end{eqnarray} 
with $\calm^2_{\rm tv}=\calm^2(\phi_{\rm tv})$.

%%%%%%%%%%%%%%%%%%%%%%%%%%%%%%%%%%%%%%%%%%%%%%%%%%%%%%%%%%%%%%%%%%%%%%%%%

\subsection{The Hartree effective potential}
\label{renormeffpothartree}
In the Hartree approximation the self-consistent
effective potential as a function of $\phi$ is obtained from a 
variational potential
\begin{eqnarray}
\calu(\phi,\calm^2)&=&U(\phi)
+\frac{1}{2}\ln\det\frac{-\partial^2+\calm^2}{-\partial+m^2}
-\frac{3\lambda}{8}\Delta^2 +\delta U_{\rm div}
+\delta U_{\rm fin}(\phi)\nonumber\\\nonumber
&=& \Lambda_{\rm fin}
+\delta \rho_{\rm fin} \phi+ \frac{m^2+\delta m^2_{\rm fin}}{2}\phi^2-
(\eta+\delta \eta_{\rm fin})
\phi^3+\frac{\lambda+\delta\lambda_{\rm fin}}{8}\phi^4
\\\nonumber &&-\frac{\calm^4}{64\pi^2}
\left(L_\epsilon-\ln\frac{\calm^2}{m^2}+\frac{3}{2}\right)
+\frac{m^4}{64\pi^2}\left(L_\epsilon+\frac{3}{2}\right)
-\frac{3\lambda}{8}\Delta^2
\\&&+\Lambda_{\rm div}
+B\calm^4\kma\label{variationalpotential}
\end{eqnarray}
with
\be\label{deltadef}
\Delta(\phi,\calm^2)=\frac{2}{3\lambda}
\left(\calm^2-m^2-\delta m^2_{\rm fin}+6(\eta+\delta\eta_{\rm fin})\phi
-\frac{3}{2}(\lambda+\delta\lambda_{\rm fin})\phi^2\right)
\kma\ee
by the condition
\be \label{extremum}
\partial \calu (\phi,\calm^2)/\partial \calm^2=0
\pkt\ee
The extremum is a maximum, so one has
\be
U_{\rm eff}(\phi)=\max_{\calm^2}~\calu(\phi,\calm^2)
\pkt\end{equation}
The condition \eqn{extremum} yields the gap equation
\bea \nonumber
\calm^2&=&m^2+\delta m^2_{\rm fin}-6(\eta+\delta \eta_{\rm fin})\phi
+\frac{3}{2}(\lambda+\delta \lambda_{\rm fin})\phi^2
\\&&+\frac{3}{2}\lambda\left[-\frac{\calm^2}{16\pi^2}
\left(L_\epsilon-\ln\frac{\calm^2}{m^2}+1\right)
+4B\calm^2\right]\kma\label{gapren}
\eea
so that
\be
\Delta=-\frac{\calm^2}{16\pi^2}\left(L_\epsilon-\ln\frac{\calm^2}{m^2}+1\right)
+4B\calm^2 
\pkt
\ee
Nevertheless, for the variation of the potential
$\calu(\phi,\calm^2)$ of Eq. \eqn{variationalpotential}, 
$\Delta$ is defined as in Eq. \eqn{deltadef}, i.e., as a
function of $\phi$ and $\calm^2$.
Cancellation of divergences in the gap equation requires
$B=L_\epsilon/64\pi^2$ up to finite terms. It is convenient 
to choose 
\be
B=\frac{L_\epsilon+1}{64\pi^2}
\pkt\ee
Cancellation of divergences in the effective action then entails
\be
\Lambda_{\rm div}=-Bm^4=-\frac{L_\epsilon+1}{64\pi^2}m^4\kma
\ee
and we  have
\be
\Delta=\frac{\calm^2}{16\pi^2}\ln\frac{\calm^2}{m^2}
\kma \ee
so that $\Delta(0,\calm^2(0))=\Delta(0,m^2)=0$.
If we want to impose, at $\phi=0$,
 the condition $\calm^2(0)=m^2$ we obtain
\be \label{constraint1}
\delta m^2_{\rm fin}=0
\pkt\ee
The condition $U_{\rm eff}(0)=0$ fixes the finite cosmological
constant to
\be
\Lambda_{\rm fin}=\frac{3\lambda}{8}\Delta^2(0,m^2)=0
\pkt\ee
We now consider the first derivative of the effective potential.
It is given by
\be
U'_{\rm eff}(\phi)=\frac{dU_{eff}(\phi)}{d \phi}=
\frac{\partial \calu(\phi,\calm^2)}{\partial \phi}
+\frac{\partial \calu(\phi,\calm^2)}{\partial \calm^2}\frac{d\calm^2}{d\phi}
 \kma\ee
with $\calm^2$ taken as the solution of the gap equation. Thereby
the second term vanishes  and so
\bea \label{firstderivative}
U'_{\rm eff}(\phi)&=&\delta\rho_{\rm fin}+m^2\phi
-3(\eta+\delta \eta_{\rm fin})\phi^2+\frac{\lambda+\delta\lambda_{\rm fin}}{2}
\phi^3
\\\nonumber
&&-\frac{1}{2}
\left[6(\eta+\delta \eta_{\rm fin})
-3(\lambda+\delta\lambda_{\rm fin})\phi\right]\Delta(\phi,\calm^2)
\pkt\eea
The requirement $U'_{\rm eff}(0)=0$, together with the condition
$\calm^2(0)=m^2$ then leads to
\be
\delta\rho_{\rm fin}=3(\eta+\delta\eta_{\rm fin})
\Delta(0,m^2)=0
\pkt\ee
With the present choice of the finite renormalizations we have
\be
U_{\rm eff}(\phi)=
 \frac{m^2}{2}\phi^2-
(\eta+\delta \eta_{\rm fin})
\phi^3+\frac{\lambda+\delta\lambda_{\rm fin}}{8}\phi^4
+\frac{\calm^4}{64\pi^2}
\left(\ln\frac{\calm^2}{m^2}-\frac{1}{2}\right)
+\frac{m^4}{128\pi^2}
-\frac{3\lambda}{8}\Delta^2
\kma \ee
and 
\be 
U'_{\rm eff}(\phi)=m^2\phi
-3(\eta+\delta \eta_{\rm fin})\phi^2+\frac{\lambda+\delta\lambda_{\rm fin}}{2}
\phi^3 -\frac{1}{2}\left[6(\eta+\delta \eta_{\rm fin})
-3(\lambda+\delta\lambda_{\rm fin})\phi\right]\Delta
\pkt\ee
The condition $U_{\rm eff}(\phi_{\rm tv})=U(\phi_{\rm tv})$
yields
\be
\delta \eta_{\rm fin}\phi_{\rm tv}^3-\frac{\delta\lambda_{\rm fin}}{8}\phi_{\rm tv}^4=
\frac{\calm^4}{64\pi^2}
\left(\ln\frac{\calm^2}{m^2}-\frac{1}{2}\right)
+\frac{m^4}{128\pi^2}
-\frac{3\lambda}{8}\Delta^2
\kma \ee
and the condition $U'_{\rm eff}(\phi_{\rm tv})=0$
leads to
\be 
3\delta \eta_{\rm fin}(\phi_{\rm tv}^2+\Delta)
-\frac{\delta\lambda_{\rm fin}}{2}\phi_{\rm tv}(\phi_{\rm tv}^2
+3\Delta)=
 -\frac{1}{2}\left[6\eta
-3\lambda\phi_{\rm tv}\right]\Delta
\kma\ee
where in both equations $\calm^2$ and $\Delta$ are taken at
$\phi=\phi_{\rm tv}$. As $\delta \eta_{\rm fin}$ and
$\delta \lambda_{\rm fin}$ appear in the equations for $\calm^2$ and
$\Delta$ we obtain a {\em nonlinear} system of equations for
$\delta\eta_{\rm fin}$ and $\delta\lambda_{\rm fin}$. 
We have solved this system  numerically using an  iterative procedure.

In the $\overline{MS}$ scheme one generally lets $L_\epsilon \to 0$ and
does not introduce any counterterms. Here we have to ensure that
we have bare vacuum conditions at $\phi=0$ with $\calm^2(0)=m^2$.
This is obtained by setting $L_\epsilon+1\to 0$ and omitting all further
counter terms. Then of course, as in the one-loop case
with $\overline{MS}$ scheme,  the minimum at $\phi_{\rm tv}$ is
shifted away from its bare value.

%%%%%%%%%%%%%%%%%%%%%%%%%%%%%%%%%%%%%%%%%%%%%%%%%%%%%%%%%%%%%%%%%%%%%%%%%

\section{Renormalization of the effective action}
\label{appendix:renormaction}
\setcounter{equation}{0}
\subsection{Renormalization of the equation of motion 1-loop approximation}
\label{renormequmot1loop}
The equation of motion for the classical profile is given by
\begin{equation}
\phi''+\frac{3}{r}\phi'-U'(\phi)+\delta U'(\phi)+\frac{1}{2}U'''(\phi)\calf=0
 \pkt\end{equation}
The fluctuation integral is decomposed into the divergent leading order
terms and a finite part as
\be
\calf(x)=\calf^{(0)}(x)+\calf^{(1)}(x)+\calf^{\overline{(2)}}(x)
\pkt\ee
The computation of the finite part $\calf^{\overline{(2)}}$
has been described in section \ref{greensfunction}.
The leading order parts are given analytically as
\begin{equation}\label{calf0}
\calf^{(0)}(x)=\int\frac{d^4k}{(2\pi)^4}\frac{1}{k^2+m^2}
=-\frac{m^2}{16\pi^2}\left(L_\epsilon+1\right)
\end{equation}
and 
\begin{eqnarray}
\calf^{(1)}(x)&=&-\int d^4y\int\frac{d^4kd^4k'}{(2\pi)^8}
\frac{e^{i(k- k')\cdot( x- y)}}
{(k^2+m^2)(k'^2+m^2)}V(y)\nonumber\\
&=&-\int\frac{d^4k}{(2\pi)^4}
\int\frac{d^4q}{(2\pi)^4}\frac{e^{iq\cdot x}\tilde{V}(q)}
{(k^2+m^2)((k+q)^2+m^2)}
\kma\end{eqnarray}
where we have defined the Fourier transformation
\begin{equation}\label{fourier}
\tilde{V}(q)=\int d^4ye^{-iq\cdot y}V(y)
\pkt\end{equation}
The integration over $k$ can be done and one obtains
\be\label{calf1}
\calf^{(1)}(x)=
-\frac{1}{16\pi^2}L_\epsilon V(x)
+\calf^{(1)}_{\rm fin}(x)
\ee
with 
\bea\calf^{(1)}_{\rm fin}(x)
&=&
-\frac{1}{16\pi^2}\int\frac{d^4q}{(2\pi)^4}e^{iq\cdot x}\tilde{V}(q)
\\\nonumber
&& \times \left(2-\frac{\sqrt{|q|^2+4m^2}}{|q|}
\ln\frac{\sqrt{|q|^2+4m^2}+|q|}{\sqrt{|q|^2+4m^2}-|q|}\right)
\pkt\eea
As the integrands, except for the exponentials, depend only on the
absolute values of $x$ and $q$ the Fourier transforms reduce to
Fourier-Bessel transform, we have
\be \label{FBV}
\tilde{V}(q)\to\tilde{V}(|q|)=
\frac{4\pi^2}{|q|}\int_0^\infty drr^2J_1(|q|r)V(r)
\pkt\ee
and similarly for $\calf^{(1)}(x)\to \calf^{(1)}(r)$.

In the equation of motion the fluctuation term and the counterterm
potential are now given by 
\begin{eqnarray}
&&\delta U'(\phi)+\frac{1}{2}U'''(\phi)\calf=\delta \rho
+\delta m^2 \phi-3\delta \eta \phi^2+\frac{1}{2}
\delta \lambda\phi^3\nonumber\\
&+&\frac{1}{2}(3\lambda\phi-6\eta)\Big[-\frac{m^2}{16\pi^2}(L_\epsilon+1)
-(\frac{3}{2}\lambda\phi^2-6\eta\phi)\frac{1}{16\pi^2}L_\epsilon
\nonumber\\
&+&\calf^{(1)}_{\rm fin}
+\calf^{\overline{(2)}}\Big]\nonumber
\pkt\end{eqnarray}
With the counterterms determined in Appendix A
the divergent terms cancel and we get the  finite expression
\bea
\delta U'(\phi)+\frac{1}{2}U'''(\phi)\calf
&=&-3\delta \eta_{\rm fin}\phi^2+\frac{1}{2}\delta \lambda_{\rm fin}
\phi^3
\\\nonumber
&&+\frac{1}{2}(3\lambda\phi-6\eta)\Big[
\calf^{(1)}_{\rm fin}+\calf^{\overline{(2)}}\Big]\nonumber
\pkt\eea

%%%%%%%%%%%%%%%%%%%%%%%%%%%%%%%%%%%%%%%%%%%%%%%%%%%%%%%%%%%%%%%%%%%%%%%%%

\subsection{Renormalization of the action in 1-loop approximation}
\label{renormaction1loop}
The 1-loop part of the effective action is given by
\begin{equation}
S_{1-l}=\frac{1}{2}\ln\cald
\pkt\end{equation}
The logarithm of the fluctuation determinant can be expanded
with respect to powers in the external potential $V(x)$ as
\be
\ln\cald =\ln\frac{-\partial^2+U''(\phi)}{-\partial^2+U''(0)}=
\sum_{N=1}^\infty\frac{(-1)^{N+1}}{N}A^N
\pkt
\ee
The first two terms in this expansion contain divergent parts.
We write
\begin{equation}
\ln\cald=A^{(1)}-\frac{1}{2}A^{(2)}+(\ln\cald)^{\overline{(3)}}
\end{equation}
and now consider the first two terms separatly.
\begin{equation}
A^{(1)}=\int\frac{d^4k}{(2\pi)^4}\frac{1}{k^2+m^2}\int d^4x V(x)
\pkt\end{equation}
Using dimensional regularization we get
\be
A^{(1)}=-\frac{m^2}{16\pi^2}L_\epsilon\int d^4x V(x)+A^{(1)}_{\rm fin}
\ee
with 
\be
A^{(1)}_{\rm fin}=-\frac{m^2}{16\pi^2}\int d^4x V(x)
\pkt
\ee
For  $A^{(2)}$ we have
\bea
A^{(2)}&=&\int\frac{d^4k}{(2\pi)^4}\frac{d^4k'}{(2\pi)^4}
\frac{\int d^4xd^4y~e^{i(k- k')\cdot(x-y)}
V(x)V(y)}
{(k^2+m^2)(k'^2+m^2)}\nonumber\\
&=&\frac{1}{16\pi^2}L_\epsilon \int d^4x(V(x))^2
+A^{(2)}_{\rm fin}
\eea
with 
\begin{equation}
A_{\rm fin}^{(2)}=
\frac{1}{128\pi^4}\int_0^\infty q^3dq|\tilde{V}(q)|^2
\left[2-\frac{\sqrt{q^2+4m^2}}{q}\ln\frac{\sqrt{q^2+4m^2}+q}
{\sqrt{q^2+4m^2}-q}\right]
\pkt\end{equation}
With  the counterterms determined in Appendix A the full
one-loop action becomes
\bea
S_{\rm eff}&=&S_{\rm cl}+S_{\rm 1l}
\\
\nonumber
&=&S_{\rm cl}+\frac{1}{2}\ln\cald^{\overline{(3)}}
+\int d^4x \left(-\delta\eta_{\rm fin}\phi^3
+\frac{\delta\lambda_{\rm fin}}{8}
\phi^4\right) -\frac{1}{4}A_{\rm fin}^{(2)}
\pkt\eea
The subtracted logarithm of the fluctuation determinant is evaluated 
according to section  \ref{flucdet}. For  
 $A_{\rm fin}^{(2)}$, as well as for $\calf^{(1)}_{\rm fin}$,
we have analytical expressions, their evaluation
involves numerical Fourier-Bessel transforms like the one
in Eq. \eqn{FBV}.

%%%%%%%%%%%%%%%%%%%%%%%%%%%%%%%%%%%%%%%%%%%%%%%%%%%%%%%%%%%%%%%%%%%%%%%%

\subsection{Renormalization in the Hartree approximation}
\label{renormhartree}

Using the counterterms of Appendix \ref{renormeffpothartree}
and the analysis of the fluctuation integral in Appendix 
\ref{renormequmot1loop}, the finite gap equation has the form :
\begin{eqnarray}\nonumber
\calm^2(x)&=&m^2-6(\eta+\delta\eta_{\rm fin})
\phi(x)+\frac{3}{2}(\lambda+\delta\lambda_{\rm fin})\phi^2(x)
\\&&+\frac{3}{2}\lambda \calf_{\rm fin}(x)\label{gapdynrenH}
\end{eqnarray}
with
\begin{equation}\label{calffinH}
\calf_{\rm fin}(x)=\frac{\calm^2-m^2}{16\pi^2}
+\calf^{(1)}(x)+\calf^{\overline{(2)}}(x)
\pkt
\end{equation}
The first term arises from $\calf^{(0)}$ in Eq. \eqn{calf0}, 
the divergent part of $\calf^{(1)}$ in Eq. \eqn{calf1} and the counterterm
$4B\calm^2=\calm^2(L_\epsilon+1)/16\pi^2$ in the
gap equation \eqn{gapren}. 
Once the profile and the fluctuation integral have been
computed Eqns. \eqn{gapdynrenH} and \eqn{calffinH} determine $\calm^2$. 
The finite equation of motion becomes
\begin{equation}
-\Delta_4\phi+U'(\phi)+
\delta U'_{\rm fin}(\phi)
-\frac{1}{2}(-6(\eta+\delta\eta_{\rm fin}+
3(\lambda+\delta\lambda_{\rm fin}\phi)\calf_{\rm fin}=0
\pkt\end{equation}
In this case the action is given by
\begin{eqnarray}
S_{\rm eff}&=&S_{\rm cl}+\delta S_{\rm fin}
+\frac{1}{2}\ln\cald^{\overline{(3)}}
-\frac{1}{4}A^{(2)}_{\rm fin}+\frac{1}{64\pi^2}\int d^4x V^2(x)
\nonumber\\
&-&\frac{3}{8}\lambda\int d^4x\calf_{\rm fin}^2
\end{eqnarray}
with
\be
\delta S_{\rm fin}=\int d^4x \left[-\delta\eta_{\rm fin}\phi^3(x)
+\frac{\delta\lambda_{\rm fin}}{8}\phi^4(x)\right]
\pkt \ee

In the $\overline{MS}$ scheme all finite renormalizations are omitted
in the gap equation, in the equation of motion and in the action,
so that the latter reduces to
\bea
S_{\rm eff}&=&S_{\rm cl}
+\frac{1}{2}\ln\cald^{\overline{(3)}}
-\frac{1}{4}A^{(2)}_{\rm fin}+\frac{1}{64\pi^2}\int d^4x V^2(x)
\nonumber\\
&-&\frac{3}{8}\lambda\int d^4x\calf_{\rm fin}^2
\pkt\end{eqnarray}
Of course the fluctuations in $\ln\cald^{\overline{(3)}}$ 
and $\calf_{\rm fin}$ and the potential $V(x)=\calm^2(x)-m^2)$ 
are computed with a
different $\calm^2$ and the profiles in  $A^{(2)}$ and in the classical
action are computed using a different equation of motion.

%%%%%%%%%%%%%%%%%%%%%%%%%%%%%%%%%%%%%%%%%%%%%%%%%%%%%%%%%%%%%%%%%%%%%%%%

\end{appendix}

%%%%%%%%%%%%%%%%%%%%%%%%%%%%%%%%%%%%%%%%%%%%%%%%%%%%%%%%%%%%%%%%%%%%%%%%
\bibliography{bounce4D}

\begin{thebibliography}{10}

\bibitem{Coleman:1977py}
S.~R. Coleman,
\newblock Phys. Rev. {\bf D15}, 2929 (1977).
%%CITATION = PHRVA,D15,2929;%%

\bibitem{Callan:1977pt}
J.~Callan, Curtis~G. and S.~R. Coleman,
\newblock Phys. Rev. {\bf D16}, 1762 (1977).
%%CITATION = PHRVA,D16,1762;%%

\bibitem{Linde:1981zj}
A.~D. Linde,
\newblock Nucl. Phys. {\bf B216}, 421 (1983).
%%CITATION = NUPHA,B216,421;%%

\bibitem{Cormier:1999ia}
D.~Cormier and R.~Holman,
\newblock Phys. Rev. {\bf D62}, 023520 (2000), [hep-ph/9912483].
%%CITATION = HEP-PH 9912483;%%

\bibitem{Cormier:1998nt}
D.~Cormier and R.~Holman,
\newblock Phys. Rev. {\bf D60}, 041301 (1999), [hep-ph/9812476].
%%CITATION = HEP-PH 9812476;%%

\bibitem{Boyanovsky:1999wd}
D.~Boyanovsky, H.~J. de~Vega and R.~Holman,
\newblock hep-ph/9903534.
%%CITATION = HEP-PH 9903534;%%

\bibitem{Garcia-Bellido:2002aj}
J.~Garcia-Bellido, M.~Garcia~Perez and A.~Gonzalez-Arroyo,
\newblock Phys. Rev. {\bf D67}, 103501 (2003), [hep-ph/0208228].
%%CITATION = HEP-PH 0208228;%%

\bibitem{Coleman:1980aw}
S.~R. Coleman and F.~De~Luccia,
\newblock Phys. Rev. {\bf D21}, 3305 (1980).
%%CITATION = PHRVA,D21,3305;%%

\bibitem{Hackworth:2004xb}
J.~C. Hackworth and E.~J. Weinberg,
\newblock Phys. Rev. {\bf D71}, 044014 (2005), [hep-th/0410142].
%%CITATION = HEP-TH 0410142;%%

\bibitem{Baacke:2003uw}
J.~Baacke and G.~Lavrelashvili,
\newblock Phys. Rev. {\bf D69}, 025009 (2004), [hep-th/0307202].
%%CITATION = HEP-TH 0307202;%%

\bibitem{Baacke:1993ne}
J.~Baacke and V.~G. Kiselev,
\newblock Phys. Rev. {\bf D48}, 5648 (1993), [hep-ph/9308273].
%%CITATION = HEP-PH 9308273;%%

\bibitem{Dunne:2005rt}
G.~V. Dunne and H.~Min,
\newblock Phys. Rev. {\bf D72}, 125004 (2005), [hep-th/0511156].
%%CITATION = HEP-TH 0511156;%%

\bibitem{Baacke:1994bk}
J.~Baacke and T.~Daiber,
\newblock Phys. Rev. {\bf D51}, 795 (1995), [hep-th/9408010].
%%CITATION = HEP-TH 9408010;%%

\bibitem{Baacke:1994ix}
J.~Baacke and S.~Junker,
\newblock Phys. Rev. {\bf D50}, 4227 (1994), [hep-th/9402078].
%%CITATION = HEP-TH 9402078;%%

\bibitem{Baacke:1993aj}
J.~Baacke and S.~Junker,
\newblock Phys. Rev. {\bf D49}, 2055 (1994), [hep-ph/9308310].
%%CITATION = HEP-PH 9308310;%%

\bibitem{Surig:1997ne}
A.~Surig,
\newblock Phys. Rev. {\bf D57}, 5049 (1998), [hep-ph/9706259].
%%CITATION = HEP-PH 9706259;%%

\bibitem{Bergner:2003id}
Y.~Bergner and L.~M.~A. Bettencourt,
\newblock Phys. Rev. {\bf D69}, 045012 (2004), [hep-ph/0308107].
%%CITATION = HEP-PH 0308107;%%

\bibitem{Baacke:2004xk}
J.~Baacke and N.~Kevlishvili,
\newblock Phys. Rev. {\bf D71}, 025008 (2005), [hep-th/0411162].
%%CITATION = HEP-TH 0411162;%%

\bibitem{Coleman85}
S.~Coleman,
\newblock {\em Aspects of Symmetry} (Cambridge University Press, 1985).

\bibitem{Coppens:1993zc}
M.~Coppens and H.~Verschelde,
\newblock Z. Phys. {\bf C58}, 319 (1993).
%%CITATION = ZEPYA,C58,319;%%

\bibitem{Verschelde:1992bs}
H.~Verschelde and M.~Coppens,
\newblock Phys. Lett. {\bf B287}, 133 (1992).
%%CITATION = PHLTA,B287,133;%%

\bibitem{Verschelde:2000dz}
H.~Verschelde,
\newblock Phys. Lett. {\bf B497}, 165 (2001), [hep-th/0009123].
%%CITATION = HEP-TH 0009123;%%

\bibitem{mottola}
E.~Mottola,
\newblock Phys. Rev. D {\bf 31}, 754 (1985).

\bibitem{bateman}
A.~Erdelyi, editor,
\newblock {\em Higher Transcendental Functions} (McGraw-Hill Book Company,
  Inc., New York, 1953).

\bibitem{Baacke:1991nh}
J.~Baacke,
\newblock Z. Phys. {\bf C53}, 402 (1992).
%%CITATION = ZEPYA,C53,402;%%

\bibitem{Baacke:1991sa}
J.~Baacke,
\newblock Acta Phys. Polon. {\bf B22}, 127 (1991).
%%CITATION = APPOA,B22,127;%%

\bibitem{Dashen:1974ci}
R.~F. Dashen, B.~Hasslacher and A.~Neveu,
\newblock Phys. Rev. {\bf D10}, 4114 (1974).
%%CITATION = PHRVA,D10,4114;%%

\bibitem{Gelfand:1959nq}
I.~M. Gelfand and A.~M. Yaglom,
\newblock J. Math. Phys. {\bf 1}, 48 (1960).
%%CITATION = JMAPA,1,48;%%

\bibitem{Verschelde:2000ta}
H.~Verschelde and J.~De~Pessemier,
\newblock Eur. Phys. J. {\bf C22}, 771 (2002), [hep-th/0009241].
%%CITATION = HEP-TH 0009241;%%

\bibitem{Nemoto:1999qf}
Y.~Nemoto, K.~Naito and M.~Oka,
\newblock Eur. Phys. J. {\bf A9}, 245 (2000), [hep-ph/9911431].
%%CITATION = HEP-PH 9911431;%%

\bibitem{Baacke:2003bt}
J.~Baacke and A.~Heinen,
\newblock Phys. Rev. {\bf D69}, 083523 (2004), [hep-ph/0311282].
%%CITATION = HEP-PH 0311282;%%

\bibitem{Dine:1992vs}
M.~Dine, R.~G. Leigh, P.~Huet, A.~D. Linde and D.~A. Linde,
\newblock Phys. Lett. {\bf B283}, 319 (1992), [hep-ph/9203201].
%%CITATION = HEP-PH 9203201;%%

\bibitem{Alford:1993zf}
M.~G. Alford, H.~Feldman and M.~Gleiser,
\newblock Phys. Rev. {\bf D47}, 2168 (1993).
%%CITATION = PHRVA,D47,2168;%%

\bibitem{Alford:1993ph}
M.~G. Alford and M.~Gleiser,
\newblock Phys. Rev. {\bf D48}, 2838 (1993), [hep-ph/9304245].
%%CITATION = HEP-PH 9304245;%%

\bibitem{Moore:2001vf}
G.~D. Moore, K.~Rummukainen and A.~Tranberg,
\newblock JHEP {\bf 04}, 017 (2001), [hep-lat/0103036].
%%CITATION = HEP-LAT 0103036;%%

\bibitem{Baacke:1998nm}
J.~Baacke and H.~Sprenger,
\newblock Phys. Rev. {\bf D60}, 054017 (1999), [hep-ph/9809428].
%%CITATION = HEP-PH 9809428;%%

\bibitem{Baacke:2000bv}
J.~Baacke and H.~Sprenger,
\newblock Phys. Rev. {\bf D63}, 094016 (2001), [hep-ph/0011204].
%%CITATION = HEP-PH 0011204;%%

\bibitem{Graham:2002fw}
N.~Graham {\em et~al.},
\newblock Phys. Lett. {\bf B572}, 196 (2003), [hep-th/0207205].
%%CITATION = HEP-TH 0207205;%%

\bibitem{Graham:2002xq}
N.~Graham {\em et~al.},
\newblock Nucl. Phys. {\bf B645}, 49 (2002), [hep-th/0207120].
%%CITATION = HEP-TH 0207120;%%

\bibitem{Graham:2002fi}
N.~Graham, R.~L. Jaffe and H.~Weigel,
\newblock Int. J. Mod. Phys. {\bf A17}, 846 (2002), [hep-th/0201148].
%%CITATION = HEP-TH 0201148;%%

\bibitem{Burnier:2005he}
Y.~Burnier and M.~Shaposhnikov,
\newblock Phys. Rev. {\bf D72}, 065011 (2005), [hep-ph/0507130].
%%CITATION = HEP-PH 0507130;%%

\bibitem{Goodband:1995qs}
M.~Goodband and M.~Hindmarsh,
\newblock Phys. Lett. {\bf B370}, 29 (1996), [hep-ph/9510434].
%%CITATION = HEP-PH 9510434;%%

\bibitem{Nagasawa:2002at}
M.~Nagasawa and R.~Brandenberger,
\newblock Phys. Rev. {\bf D67}, 043504 (2003), [hep-ph/0207246].
%%CITATION = HEP-PH 0207246;%%

\bibitem{Bordag:2002sa}
M.~Bordag,
\newblock Phys. Rev. {\bf D67}, 065001 (2003), [hep-th/0211080].
%%CITATION = HEP-TH 0211080;%%

\bibitem{Diakonov:1987ty}
D.~Diakonov, V.~Y. Petrov and P.~V. Pobylitsa,
\newblock Nucl. Phys. {\bf B306}, 809 (1988).
%%CITATION = NUPHA,B306,809;%%

\bibitem{Christov:1995vm}
C.~V. Christov {\em et~al.},
\newblock Prog. Part. Nucl. Phys. {\bf 37}, 91 (1996), [hep-ph/9604441].
%%CITATION = HEP-PH 9604441;%%

\bibitem{Alkofer:1994ph}
R.~Alkofer, H.~Reinhardt and H.~Weigel,
\newblock Phys. Rept. {\bf 265}, 139 (1996), [hep-ph/9501213].
%%CITATION = HEP-PH 9501213;%%

\end{thebibliography}
\bibliographystyle{h-physrev4}

\end{document}